%% file: main.tex
\documentclass[twocolumn, times, trackchanges]{aastex63}

\graphicspath{{./}{figures/}}
\turnoffedit

\newcommand{\RomanNumeralCaps}[1]
    {\MakeUppercase{\romannumeral #1}}

\begin{document}

\title{Efficiently Searching for Close-in Companions around Young M Dwarfs using a Multi-year PSF Library}

\correspondingauthor{Aniket Sanghi}
\email{asanghi@caltech.edu}

\author[0000-0002-1838-4757]{Aniket Sanghi}
\altaffiliation{NSF Graduate Research Fellow.}
\affiliation{Cahill Center for Astronomy and Astrophysics, California Institute of Technology, 1200 E. California Boulevard, MC 249-17, Pasadena, CA 91125, USA}

\author[0000-0002-6618-1137]{Jerry W. Xuan}
\affiliation{Cahill Center for Astronomy and Astrophysics, California Institute of Technology, 1200 E. California Boulevard, MC 249-17, Pasadena, CA 91125, USA}

\author[0000-0003-0774-6502]{Jason J. Wang}
\affiliation{Center for Interdisciplinary Exploration and Research in Astrophysics (CIERA) and Department of Physics and Astronomy, Northwestern University, Evanston, IL 60208, USA}
\affiliation{Cahill Center for Astronomy and Astrophysics, California Institute of Technology, 1200 E. California Boulevard, MC 249-17, Pasadena, CA 91125, USA}

\author[0000-0002-8895-4735]{Dimitri Mawet}
\affiliation{Cahill Center for Astronomy and Astrophysics, California Institute of Technology, 1200 E. California Boulevard, MC 249-17, Pasadena, CA 91125, USA}

\author[0000-0003-2649-2288]{Brendan P. Bowler}
\affiliation{The University of Texas at Austin, Department of Astronomy, 2515 Speedway, C1400, Austin, TX 78712, USA}

\author[0000-0001-5172-4859]{Henry Ngo}
\affiliation{NRC Herzberg Astronomy and Astrophysics, 5071 W Saanich Road, Victoria, BC V9E 2E7, Canada}

\author[0000-0002-6076-5967]{Marta L. Bryan}
\affiliation{David A. Dunlap Institute Department of Astronomy \& Astrophysics, University of Toronto, 50 St. George Street, Toronto, ON M5S 3H4, Canada}

\author[0000-0003-4769-1665]{Garreth Ruane}
\affiliation{Jet Propulsion Laboratory, California Institute of Technology, 4800 Oak Grove Dr., Pasadena, CA 91109, USA}

\author[0000-0002-4006-6237]{Olivier Absil}
\affiliation{Universit\'e de Li\`ege, STAR Institute, All\'ee du Six Ao\^ut 19c, B-4000 Li\`ege, Belgium}

\author[0000-0002-1342-2822]{Elsa Huby}
\affiliation{LESIA, Observatoire de Paris, Universit\'e PSL, CNRS, Sorbonne Universit\'e, Universit\'e Paris Cité, 5 place Jules Janssen, 92195 Meudon, France}

\shorttitle{Keck/NIRC2 Coronagraphic Imaging Search for Close-in Companions around Young M Dwarfs}

\shortauthors{Sanghi et al.}

\begin{abstract}
We present Super-RDI, a unique framework for the application of reference star differential imaging (RDI) to Keck/NIRC2 high-contrast imaging observations with the vortex coronagraph. Super-RDI combines frame selection and signal-to-noise ratio (S/N) optimization techniques with a large multi-year reference point spread function (PSF) library to achieve optimal PSF subtraction at small angular separations. We compile a $\sim$7000 frame reference PSF library based on a set of 288 new Keck/NIRC2 $L'$ sequences of 237 unique targets acquired between 2015 and 2019 as part of two planet-search programs designed for RDI, one focusing on nearby young M dwarfs and the other targeting members of the Taurus star-forming region. For our dataset, synthetic companion injection-recovery tests reveal that frame selection with the mean-squared error (MSE) metric combined with Kahrunen Loève Image Processing-based (KLIP) PSF subtraction using 1000--3000 frames and $<$500 principal components yields the highest average S/N for injected synthetic companions. We uniformly reduce targets in the young M-star survey with both Super-RDI and angular differential imaging (ADI). For the typical parallactic angle rotation of our dataset ($\sim$10$^\circ$), Super-RDI performs better than a widely used implementation of ADI-based PSF subtraction at separations $\lesssim 0\farcs4$ ($\approx$5 $\lambda$/$D$) gaining an average of 0.25\,mag in contrast at $0\farcs25$ and 0.4\,mag in contrast at $0\farcs15$. This represents a performance improvement in separation space over RDI with single-night reference star observations ($\sim$100 frame PSF libraries) applied to a similar Keck/NIRC2 dataset in previous work. We recover two known brown dwarf companions and provide detection limits for 155 targets in the young M-star survey. Our results demonstrate that increasing the PSF library size with careful selection of reference frames can improve the performance of RDI with the Keck/NIRC2 vortex coronagraph in $L'$. 
\end{abstract}


\section{Introduction}
\label{sec:intro}
The past two decades have witnessed tremendous evolution in efforts to directly detect and characterize exoplanetary systems. Pioneering work has paired instruments such as high-order adaptive optics (AO) systems and small inner-working angle coronagraphs with innovative observing and post-processing strategies to eliminate contaminating starlight and achieve high contrast ratios \citep[e.g.,][]{2006ApJS..167...81G,2009ARA&A..47..253O, 2010Sci...329...57L,2010SPIE.7736E..1JM,2010A&ARv..18..317A, 2012SPIE.8442E..04M, 2015Sci...350...64M,2016PASP..128j2001B, 2023ASPC..534..799C}. The results revealed a previously unseen population of giant planets ($\gtrsim$1 $M_\mathrm{{Jup}}$) at wide orbital separations ($\gtrsim$10 au). Uncovering giant exoplanets at large separations has challenged theories of planet formation and evolution leading to the exploration of different mechanisms such as core accretion, dynamical scattering, disk instability, and cloud fragmentation; all of which act over different orbital distance regimes \citep[e.g.,][]{1997Sci...276.1836B, 2014prpl.conf..643H, 2014prpl.conf..619C, 2014Life....4..142H, 2016ARA&A..54..271K}. A primary challenge in investigating these has been the modest sample of directly imaged planets available for precise characterization studies.

Large demographic surveys such as the Gemini Deep Planet Survey \citep[GDPS:][]{2007ApJ...670.1367L}, Strategic Exploration of Exoplanets and Disks with Subaru \citep[SEEDS:][]{2009AIPC.1158...11T, 2016PJAB...92...45T}, the Gemini NICI Planet-finding Campaign \citep{2010SPIE.7736E..1KL}, the International Deep Planet Search \citep[IDPS:][]{2012A&A...544A...9V}, Planets Around Low Mass Stars \citep[PALMS:][]{2015ApJS..216....7B}, LBTI Exozodi Exoplanet Common Hunt \citep[LEECH:][]{2014SPIE.9148E..0LS, 2018AJ....156..286S}, the Gemini Planet Imager Exoplanet Survey \citep[GPIES:][]{2018SPIE10703E..0KM}, and the SpHere INfrared Exoplanets Survey \citep[SHINE:][]{2021A&A...651A..70D} have established the rarity of giant planets at wide separations, with occurrence rates as low as a few percent \citep[e.g.,][]{2016PASP..128j2001B, 2019AJ....158...13N, 2021A&A...651A..72V}. In contrast, radial velocity (RV) surveys uncover a peak in the giant planet occurrence rate distribution at 2--8 au \citep{2019ApJ...874...81F, 2021ApJS..255...14F}. Thus, imaging typical giant planets in the 1--10 au range beyond the water ice line but within the region where protoplanetary disk surface densities quickly fall off can eventually provide a large enough sample to carry out population-level orbital and atmospheric characterization studies. This requires sensitivity gains to be made with both instrumentation and post-processing strategies. In particular, in this study we focus on optimizing $L'$ observations with the Keck/NIRC2 vortex coronagraph.

The Keck/NIRC2 vortex coronagraph \citep{2005OptL...30.3308F, Mawet_2005} is an instrument mode that enables infrared high-contrast imaging in $L'$ and $M_s$ bands ($\lambda = 3.4$--$4.8\;\mu$m) at small angular separations from the star ($\sim$100 mas). For the nearest young stars, this capability provides unique opportunities for the detection of $<$10 au self-luminous giant planets when combined with the right observing strategy \citep[e.g., AF Lep b,][]{2023ApJ...950L..19F}. Observations with the Keck/NIRC2 vortex coronagraph have primarily employed the angular differential imaging \citep[ADI:][]{2004Sci...305.1442L, 2006ApJ...641..556M} strategy. Ground-based ADI observations are conducted with the telescope de-rotator turned off (for Cassegrain focus) or with the field rotator set to track the telescope pupil. As the field-of-view rotates under sidereal motion, astrophysical signals (e.g., companions and disk features) revolve in the image frame while the point spread function (PSF) of the host star remains stationary on the detector. Subtracting image frames obtained at different parallactic angles (P.A.s) removes the host starlight without significant subtraction of the planet signal. While ADI has proved to be a powerful observational strategy for directly imaging exoplanets, it possesses certain inherent limitations. Most prominently, observations of close-in companions suffer from self-subtraction of the planet signal. This effect can be reduced by observing the target with sufficient P.A.~rotation and employing a rotation gap criterion in post-processing (generally excluding reference frames with $\lesssim 1\; \lambda/D$ field rotation in relation to a given science frame). Such observations are difficult to schedule for a large survey and offer limited sky coverage \citep[see Figure 1 in ][]{2019AJ....157..118R}.

\begin{figure*}
    \centering
    \includegraphics[scale=0.08]{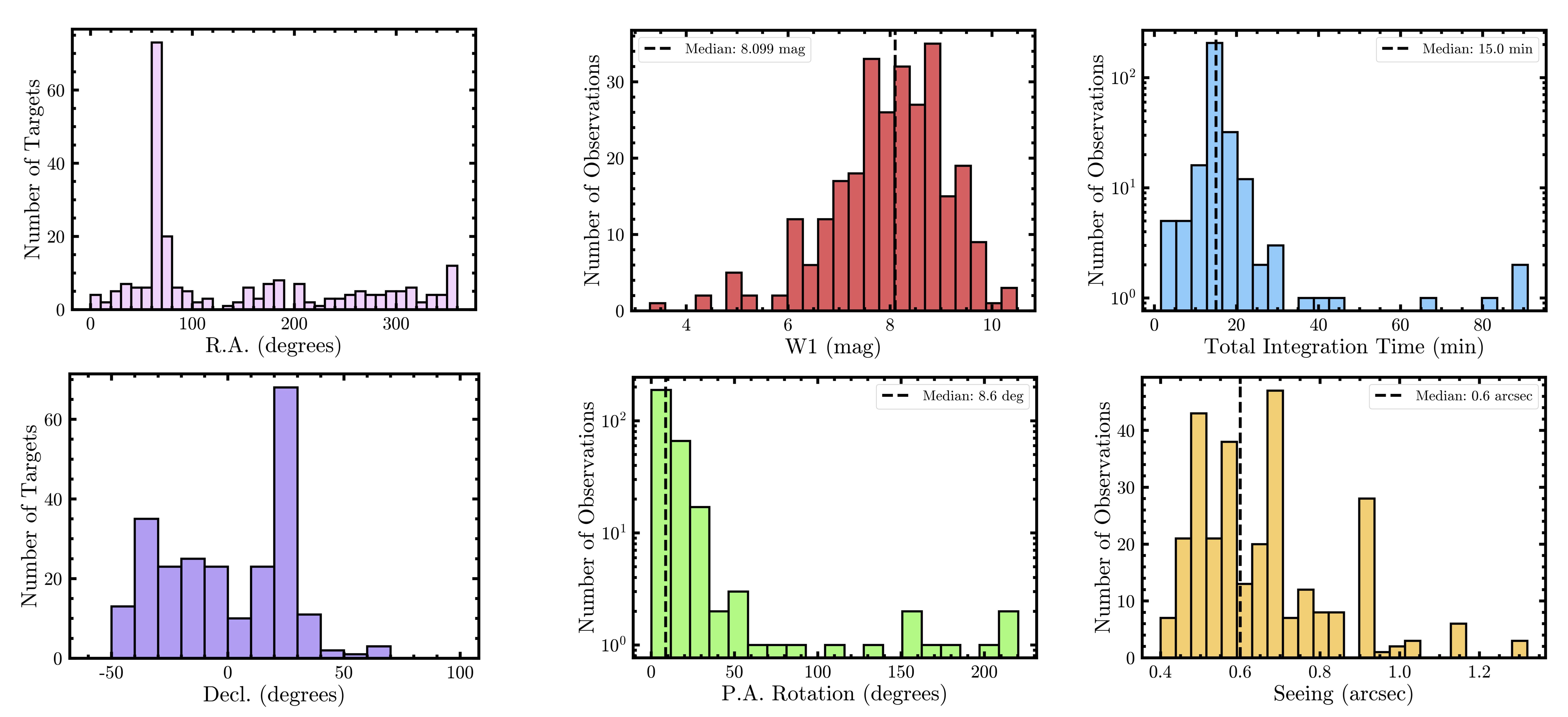}
    \caption{Histograms summarizing distributions for key properties of targets in the sample (R.A.~and Decl.) and individual observations (\emph{W1} mag, total integration time, P.A.~rotation, and seeing). The median values for the observations are as follows: 8.099\,mag \emph{W1}, 15 min of total integration time, 8.6$^\circ$ P.A. rotation, and 0\farcs6 seeing.}
    \label{fig:obs}
\end{figure*}

Reference star differential imaging \citep[RDI:][]{1984Sci...226.1421S, 2009A&A...493L..21L, Lafreni_re_2009} is one of the primary observing strategies that enables direct imaging of point sources and disks at small angular separations without the limitations posed by ADI. This approach uses reference stars distinct from the science target to build a model of the target star's PSF for primary-star subtraction. RDI performs best when the reference frames exhibit minimal deviations from the ideal PSF. Space-based applications of RDI \citep[e.g.,][]{2014ApJ...786L..23S, 2014AJ....148...59S, 2016ApJ...817L...2C, 2019JATIS...5c5003D, 2021MNRAS.504.3074W, 2022AJ....163..119S, 2023ApJ...951L..20C, 2023ApJ...945..126G} benefit from more stable conditions since they are not hindered by turbulence from the Earth's atmosphere. For ground-based observations, one way previous work has addressed the issue of varying observing conditions is by imaging designated reference stars directly before and after the science observations to ensure maximum PSF correlation \citep[e.g.,][]{2017AJ....154...73R}. When designated reference stars are unavailable, \citet{2018AJ....156..156X} found that restricting the PSF library to all reference frames imaged on the same night as the science target improved sensitivity to point sources at small angular separations ($\lesssim 0\farcs25$) with respect to ADI. \citet{2019AJ....157..118R} improve on this strategy by introducing image similarity metric-based pre-frame selection techniques, which increased detection significance by a factor of $\sim$3. Imaging binary stars \citep[][]{2015ApJ...811..157R, 2022MNRAS.515.4487P} or observing the reference star near simultaneously as the science target \citep[][]{2021A&A...648A..26W} are alternative ways of improving RDI's performance in ground-based applications. The various approaches have been tested with observations across a range of wavelengths and it may be the case that the optimal strategy is wavelength dependent.

The impact of PSF variations can be mitigated by using a diversity of reference frames to accurately create a model of the target star's PSF. Several studies have used archival PSF observations across multiple nights to compile reference libraries \citep[e.g.,][]{2012SPIE.8447E..0OM, 2016SPIE.9909E..58G, 2018MNRAS.473.1774L, 2020AJ....159..251D}. Common selection techniques involve excluding close-binaries, matching the spectral filter or observing mode between the science and reference star observations, and choosing reference frames with high correlation coefficients with the science datasets. However, most of these studies implemented these techniques for the analysis of a single science target and did not experiment with changing the size of the reference library or using multiple pre-frame selection techniques (primarily because the science focus was different). Such analyses are now important for the complete characterization of RDI's strengths and weaknesses with ground-based instruments. Recently, \citet{2022A&A...666A..32X} used all available archival VLT SPHERE/IRDIS \emph{H23} data (1000+ targets) observed over a period of five years to compile a $\sim$$7 \times 10^4$ frame reference library (after removal of poor reference stars). They analyzed the average performance of RDI across 32 representative targets in their sample as a function of seeing condition and P.A.~rotation. For observations obtained under the median seeing condition of their reference library, P.A.~rotations of $\approx$30$^\circ$, and frame selection with the mean-squared error (MSE) metric, they found that RDI outperformed ADI at separations $\lesssim 0\farcs4$. For worse seeing conditions or larger P.A.~rotations, ADI performed at a level comparable to or better than RDI. \citet{2022A&A...666A..32X} experimented with different reference library sizes (up to $10^4$ frames) and found that increasing the reference library size to 3000--5000 frames improved RDI's sensitivity level, after which it plateaued. These results motivate investigations into the performance of RDI using large reference libraries with more ground-based high-contrast imaging instruments. Specifically, Keck/NIRC2 vortex coronagraphic observations are well-suited to such a study given the large number of observations available in the archive\footnote{\url{https://koa.ipac.caltech.edu}}.

In this work, we present the Super-RDI framework, which is designed to improve the performance of the RDI strategy when working with large reference libraries. We develop and apply this framework in the context of Keck/NIRC2 high-contrast $L'$ imaging observations with the vortex coronagraph obtained as part of two high-contrast imaging surveys, the young M-star survey---which focused on young nearby M dwarfs---and the Taurus survey---which targeted predominantly low-mass members of the Taurus star-forming region\footnote{Targets were selected from compilations in \citet{2006ApJ...647.1180L, 2011ApJ...731....8K, 2014ApJ...784..126E, 2015ApJ...799..155D, 2017ApJ...838..150K}.}. Note that the Taurus survey targets only contribute to our reference library for PSF subtraction of the young M-star survey targets. PSF subtraction results of the Taurus survey targets will be presented in Bryan et al.~(in preparation). This paper is organized as follows. Section~\ref{sec:survey} describes the design of the young M-star survey from the perspective of target selection and observing strategy. Section~\ref{sec:obs} details our observations and summarizes the properties of the sample used in our analysis. Section~\ref{sec:srdi} introduces the Super-RDI framework and the steps involved in its implementation. Section~\ref{sec:adi} outlines the ADI reduction of targets in the young M-star survey sample. Section~\ref{sec:char} discusses the sensitivity achieved by Super-RDI and compares its performance to a widely used implementation of ADI-based PSF subtraction in the context of the young M-star survey sample. Section~\ref{sec:det} describes point source detections in the Super-RDI processed dataset. Finally, we summarize our conclusions in Section~\ref{sec:con}. Appendix~\ref{app} provides information about the outcomes of parameter optimization with ADI for our survey dataset. Appendix~\ref{app:B} provides optimal contrasts at select separations for targets in the young M-star survey.

\section{Young M-star Survey Design}
\label{sec:survey}
\subsection{Target Selection} \label{sec:sample}

Targets in this survey comprise young- and intermediate-age ($\approx$20--200~Myr) nearby M dwarfs within $\approx$100~pc. Precise age-dating of active M dwarfs that are not members of well-recognized young moving groups or star-forming regions can be challenging, so we defer a detailed discussion of ages to a future study. Instead, except for systems with confirmed or candidate companions, the focus of this work is on the RDI processing of these observations and the generation of contrast curves rather than the corresponding mass limits (which requires host star ages). Below we discuss details of the sample selection.

The goal of the young M-star survey was to directly build upon the first-generation of direct imaging surveys that probed the outskirts (10--100 au) of young M dwarf systems.  In particular, the Planets Around Low-Mass Stars (PALMS) survey carried out deep observations of 122 young M dwarfs with Keck/NIRC2 and Subaru/HiCIAO in $H$ and $K_S$ bands \citep{2015ApJS..216....7B}.  Several widely separated brown dwarf companions were discovered;  however, observations from this program made use of Lyot coronagraphs so sensitivity was greatly diminished at closer inner working angles. Other first-generation high-contrast imaging surveys focusing on young M dwarfs were typically smaller in size \citep[e.g.,][]{2007ApJ...654..558D, 2012A&A...539A..72D, 2016A&A...596A..83L} or were sensitive to substellar companions at even wider separations \citep[e.g.,][]{2008AJ....135.2024A, 2017AJ....154..129N}.

An expanded phase of the PALMS survey was carried out with shorter integration times of a few minutes per target to efficiently search for substellar companions around several hundred newly identified young M dwarfs (Bowler et al., in preparation). These targets were selected in the pre-Gaia era and represent a compilation of new confirmed and candidate low-mass members of young moving groups drawn from a wide range of sources, including published studies---in particular \citet{2008hsf2.book..757T, 2012AJ....143...80S, 2012ApJ...758...56S, 2013ApJ...762...88M, 2014ApJ...788...81M, 2014AJ....147..146K, 2015ApJS..219...33G, 2016A&A...590A..13E}---and a large ongoing search for new young K- and M-type members of moving groups, some of which are presented in \citet{2019ApJ...877...60B}. As part of this effort, the companions 2MASS~J01225093-2439505~B \citep{2013ApJ...774...55B}, 2MASS~J02155892--0929121~C \citep{2015ApJ...806...62B}, and 2MASS~J22362452+4751425~b \citep{2017AJ....153...18B} were discovered, which span the brown dwarf to high-mass planetary regimes.  Many close binaries with separations ranging from the Keck diffraction limit (about 50 mas in $K$ band) to several arcseconds were also found in this expanded PALMS adaptive optics imaging survey.  Among the apparently single stars, those which were the closest, youngest, and most securely linked to known young moving groups were selected for deeper imaging to probe smaller separations at $L'$-band with the newly installed NIRC2 vortex coronagraph. These are the observations we present in this study.

\subsection{Observing Strategy}
\label{sec:obs-strat}
To efficiently survey a sample of $\approx$150--200 science targets in a time-effective manner, several important factors had to be considered when selecting an optimal observing strategy. The two primary strategies available to the survey were ADI and RDI. For ADI, P.A.~rotation accumulated across an observation sequence is the key driver of achievable contrast. For RDI, similarity between the reference and science frames across variable observing conditions is the primary factor that determines achievable contrast. For this particular survey, we noted the following:
\begin{enumerate}
    \item Since one of the science goals of the survey was to better understand the demographics of giant planets and brown dwarfs around M dwarfs, an observational set-up that remained approximately consistent between targets was preferred. 
    \item A significant number of targets in our sample overlapped in R.A.~(Figure \ref{fig:obs}), implying that for a fixed number of observing nights, not all targets could be observed at the time of transit (when the P.A.~rotation is maximum).
    \item To accrue sufficient P.A.~rotation (generally $>$30$^\circ$) for ADI, targets would have to be observed for $\sim$1--2 hours \citep[e.g.,][]{2017AJ....154...73R}. Assuming 1-hour (resp.~2-hour) observing blocks, the cost for $\sim$200 targets including overheads amounts to $\sim$200 (resp.~$\sim$400) hours spread over more than 100 individual nights, if all targets are to be observed during transit (Figure \ref{fig:obs}).
\end{enumerate}

Thus, to balance efficiency and performance, we chose to primarily design the survey to leverage RDI. Since the targets in the survey are similar in spectral type, science targets observed on the same night could serve as reference stars for each other \citep[the approach explored in][]{2018AJ....156..156X} or the entire set of observations after survey completion could be used as a PSF library for RDI (the approach explored in this work). While we still conducted observations in vertical angle (ADI) mode to ensure PSF consistency, we did not accrue significant P.A.~rotation for the majority of targets in our sample (Figure \ref{fig:obs}) and thus do not expect ADI to perform at a comparable level for our survey.

\input{tab1}

\section{Observations}
\label{sec:obs}
This work's sample consists of a set of 288 sequences of 237 unique targets observed from 2015~December~26 to 2019~January~9. The sample is comprised of targets that have been observed independently as part of two survey programs with Keck/NIRC2: the young M-star survey (195 observations of 157 unique targets) and the Taurus survey (93 observations of 80 unique targets). Both sets of observations were taken with the vector vortex coronagraph, installed in the Keck/NIRC2 camera \citep{2017AJ....153...43S}, using the QACITS automatic, real-time coronagraphic PSF centering algorithm \citep{2015A&A...584A..74H, 2017A&A...600A..46H}. Typically, the centering accuracy provided by QACITS is 2.4 mas rms \citep{2017A&A...600A..46H}, or 0.025 $\lambda$/D rms in $L'$ band. In comparison, the pixel scale of the NIRC2 vortex coronagraph is 9.971 mas per pixel \citep{2016PASP..128i5004S}. The full data set contains images taken in the $L'$ (central wavelength of 3.776 $\mu$m) bandpass. For our sample, the median and mean on-source integration times (number of science frames $\times$ coadds $\times$ exposure time per coadd) are 15.0 minutes and 16.5 minutes, respectively. The median and mean seeing conditions were 0$\farcs$6 and 0$\farcs$7. Seeing is obtained from Maunakea Weather Center's\footnote{\url{http://mkwc.ifa.hawaii.edu/current/seeing/}} DIMM Seeing Monitor (mean value) or the CFHT WX Tower Seeing Monitors (mean value) if the former is unavailable. Seeing conditions were not available for a total of 27 targets observed across 4 nights: 2015~December~26, 2015~December~27, 2017~December~24, and 2018~October~21. Important properties of the sample are presented individually for each target in Table~\ref{tab1} and as distributions in Figure~\ref{fig:obs}.

The general observing sequence for targets in our full sample is summarized below. We acquire one image of the star without the coronagraph to characterize the unocculted PSF, two sky frames of a blank field 10$\arcsec$ away from the target, and then $\approx$18--40 science frames with the star centered on the vortex, representing an individual frame integration time of $\approx$0.5--1.0 s. Longer observations or those undertaken in rapidly changing conditions require the full sequence to be repeated every 10--30 minutes. This allows for sampling of potential variations in the unocculted PSF and sky background. All observations were taken with the telescope's field rotator set to track the telescope pupil in order to exploit the natural rotation of the sky for ADI. For our dataset, the median and mean P.A. rotations are 8.6$^\circ$ and 16.4$^\circ$, respectively. The range of P.A. rotations is 0.1--220.3$^\circ$.
 
The raw dataset is uniformly reprocessed using a pipeline that automatically downloads, sorts, and processes data \citep{2018AJ....156..156X}. This pipeline is based on functions available as part of the Vortex Image Processing (\texttt{VIP}) software package \citep{2017AJ....154....7G, Christiaens2023} as well as custom programs. Here, the key points from \citet{2018AJ....156..156X} are summarized below. (1) The science and sky background frames are flat field corrected using the median of 5-10 blank sky images acquired with the vortex mask removed near the end of the same or a close in time night. (2) The pipeline is used to replace the value of bad, hot, and dead pixels in the science, sky, dark and flat field frames with the median of the neighboring pixels in a $5 \times 5$ pixels box. However, we avoid bad pixel correction in a circular region of diameter equal to the full-width-at-half-maximum (FWHM) centered on the star. (3) A principal component analysis (PCA)-based algorithm subtracts the sky from the science frames (the central 1 FWHM is masked). The sky-subtracted images are registered to the target star's location and de-rotated to align north up and east left. We provide the pre-processsed coronagraphic frames, corresponding parallactic angles, and unocculted stellar PSF frame for all targets in the young M-star survey on Zenodo\footnote{\url{https://zenodo.org/records/12747613}} for public access.

\begin{figure*}
    \centering
    \includegraphics[scale=0.10]{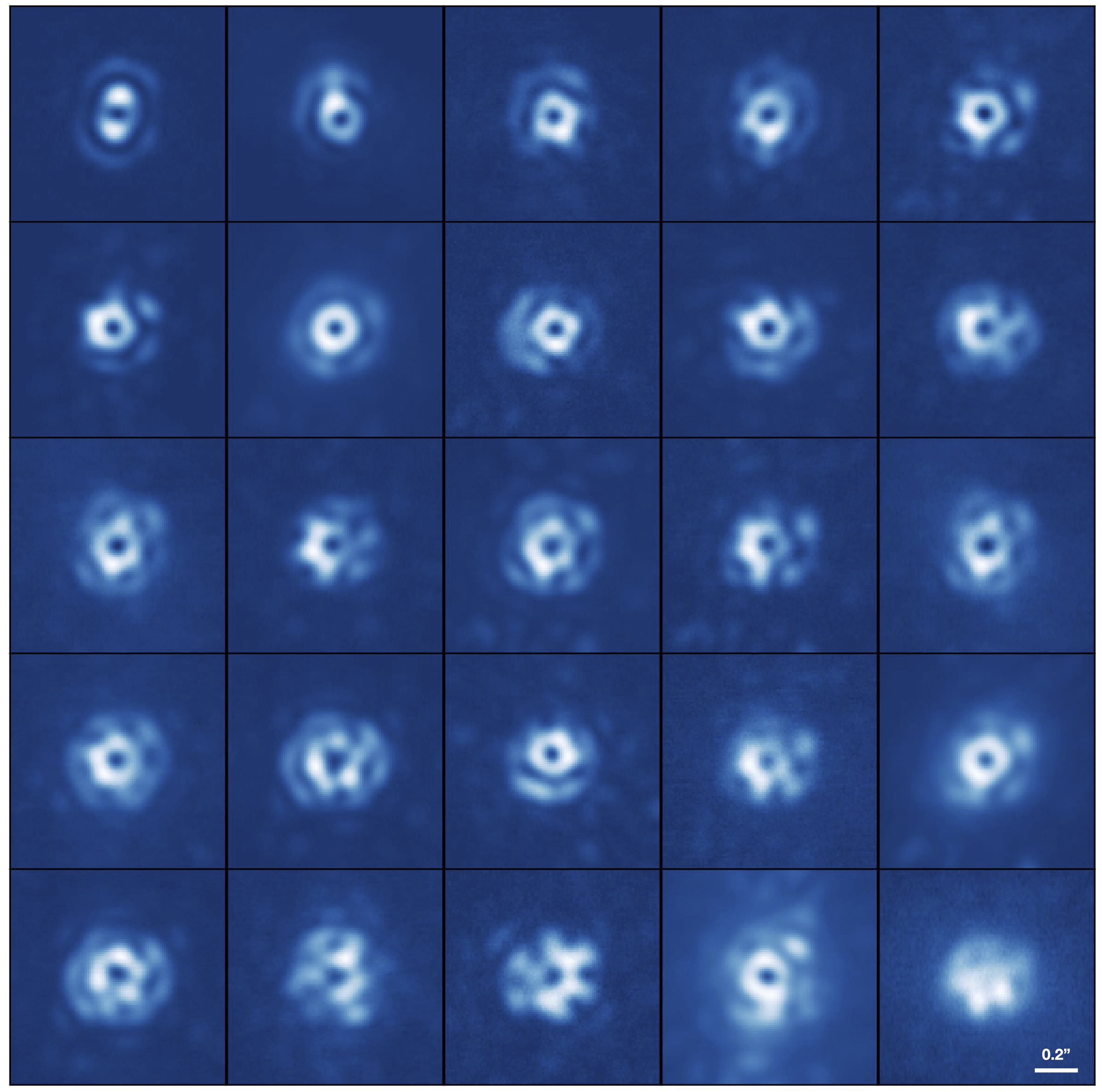}
    \caption{Gallery of 25 example sky-subtracted image cutouts (centered on target) drawn randomly from our NIRC2 vortex coronagraphic observations. The dimensions of each cutout is 101 $\times$ 101 pixels (1\arcsec $\times$ 1\arcsec). The images are depicted in linear scale in normalized analog to digital units and are ordered left to right, top to bottom in ascending order of raw contrast level at a fixed radial separation ($\sim$10$^{-2}$ at 0\farcs15). The diversity of PSF shapes highlight the importance of using frame selection techniques to achieve optimal performance with the RDI strategy.}
    \label{fig:frames}
\end{figure*}

\section{The Super-RDI Framework}
\label{sec:srdi}
In this section, we develop a multi-step framework called Super-RDI for the systematic use of a large PSF library to achieve optimal PSF subtraction at small angular separations with the RDI strategy. The framework is applicable to direct imaging data with any instrument and is not restricted to Keck/NIRC2 observations. Broadly, the framework encapsulates the following three steps: (1) use of image similarity metrics to rank and select the best matching reference stars uniquely for each target; (2) optimization of free parameters using synthetic companion injection-recovery tests to maximize detection sensitivity; and (3) uniform processing and sensitivity analysis of the dataset using the optimized set of reduction parameters. In this work, we will present the reductions and sensitivity curves of targets in the young M-star survey sample (195 observations of 157 unique targets). While the Taurus survey targets have been used as reference stars in subsequent procedures, their reductions and sensitivity curves will be presented in Bryan et al.~(in preparation). We discuss each of the above steps and their application to our sample in the following subsections. 

\subsection{Image Similarity Metrics}

\label{sec:stats}
An important factor affecting the performance of RDI is the quality of individual frames (see example gallery of PSFs in Figure~\ref{fig:frames}) in the reference library \citep{2019AJ....157..118R, 2022AJ....163..119S}. Pre-selecting reference frames is a necessary step to achieve high contrast ratios at small angular separations. This has been demonstrated in the past by several ground- and space-based high contrast imaging studies (see \S\ref{sec:intro}). A powerful technique for frame pre-selection is to use statistical image similarity metrics. Image similarity metrics aid the identification of a set of reference PSFs that are most similar, in structure and intensity distribution, to the science observations. This reduces random speckle noise in post-processed images and enables us to achieve higher sensitivity to potential companions and disks of interest. In this study, we use a set of five image similarity metrics to rank reference frames for each individual target in the sample. Three of the metrics are drawn from the study conducted by \citet{2019AJ....157..118R}: mean-squared error (MSE), Pearson's correlation coefficient (PCC), and structural similarity index metric (SSIM). The remaining two metrics have been newly developed as part of this work: flux logarithmic standard deviation indicator (FLSI) and contrast logarithmic standard deviation indicator (CLSI). 

\subsubsection{MSE, PCC, and SSIM}
We refer the reader to \citet{2019AJ....157..118R} for a detailed discussion on the MSE, PCC, and SSIM metrics as well as their quantitative definitions. Qualitative interpretations of the three metrics are described below. MSE is a measure of the absolute error between two images. A lower MSE is preferred during reference frame pre-selection. One drawback of the metric is that its value is dependent on image intensity. Thus, two stars with similar PSF structure and quality but different brightnesses may yield a high MSE value\footnote{A possible solution is to weight each reference PSF by a scaling factor that minimizes the residuals with the science PSF before calculating a MSE value. While we did not implement this, the complementarity of other metrics used in this work (such as PCC, which would yield a favorable value in this case) and that the majority of targets in our sample are of the same spectral type (M dwarfs) limit this drawback. Additionally, such a situation has a low probability of occurrence given that the PSF morphology will likely change between stars that differ in brightness due to a difference in AO performance.}. PCC is a normalized measure of correlation between two image frames. A higher PCC is preferred during reference frame pre-selection. SSIM is a perception-based model that considers changes in structural information. Structural information encapsulates the idea that pixels have strong inter-dependencies when they are spatially close. SSIM consists of a luminance, contrast, and structural term. The luminance and structural terms behave similar to the MSE and PCC respectively. Thus, SSIM can be thought of as a mixture between the MSE and PCC metrics. A higher SSIM is preferred during reference frame pre-selection.

\subsubsection{FLSI and CLSI}
We present the calculation of two new image similarity metrics that help quantify the similarity in radial flux and raw contrast distributions between two image frames. FLSI is derived as follows. Given a reference frame, we construct 1 FWHM ($\approx$8 pixels) diameter apertures at a fixed angular separation from the star center. The pixel intensities in each aperture are integrated and the mean integrated aperture flux value is determined at the given separation for each frame. This procedure is repeated for a range of separations from the center of the star to obtain the mean integrated aperture flux value as a function of separation. We term the above as the mean flux curve $F_{ik}$ for the $i$th frame of the $k$th reference star. Mean flux curves are computed for all image frames of all reference stars in the library. Next, the procedure is repeated for all frames of the science target. We average the mean flux curves corresponding to all science target frames to obtain the representative mean flux curve for the science target ($F_M$). Finally, FLSI for the $i$th frame of the $k$th reference star with respect to the science target is given by the following calculation,
\begin{equation}
    \mathrm{FLSI}_{ik} = \mathrm{std}(\mathrm{log_{10}}F_{ik} - \mathrm{log_{10}}F_M),
\end{equation}
\noindent where the standard deviation of quantity $X$ (with $N$ values and mean $\bar{X}$) is defined as,
\begin{equation}
    \mathrm{std}(X) = \sqrt{\frac{1}{N - 1}\sum_{i=1}^{N}(X_i - \bar{X})^2},
\end{equation}

\noindent This metric quantifies image similarity as similarity in radial flux distributions. While the absolute value of the flux (as characterized by mean flux curves) may differ between the science target and the chosen reference frame, a constant logarithmic difference in the mean flux curves across multiple separations would point to a good match. Mathematically, a constant logarithmic difference would be indicated by a low standard deviation in the differences measured as a function of separation (i.e., low FLSI).

We follow a similar procedure to calculate CLSI. The key difference is that instead of calculating the mean integrated aperture flux as a function of separation, we determine the 5$\sigma$ raw contrast as a function of separation. The 5$\sigma$ raw contrast is defined as five times the standard deviation of the integrated aperture fluxes at a given separation normalized by the host star flux. If $C_{ik}$ is the raw contrast curve for the $i$th frame of the $k$th reference star and $C_M$ is the average of the raw contrast curves of all frames of the science target, then CLSI for the given reference frame is calculated as
\begin{equation}
    \mathrm{CLSI}_{ik} = \mathrm{std}(\mathrm{log_{10}}C_{ik} - \mathrm{log_{10}}C_M).
\end{equation}
\noindent This metric quantifies image similarity as similarity in radial raw contrast distributions. Similar to FLSI, a low CLSI value is preferred during reference frame pre-selection.

\begin{figure}
    \centering
    \includegraphics[scale=0.083]{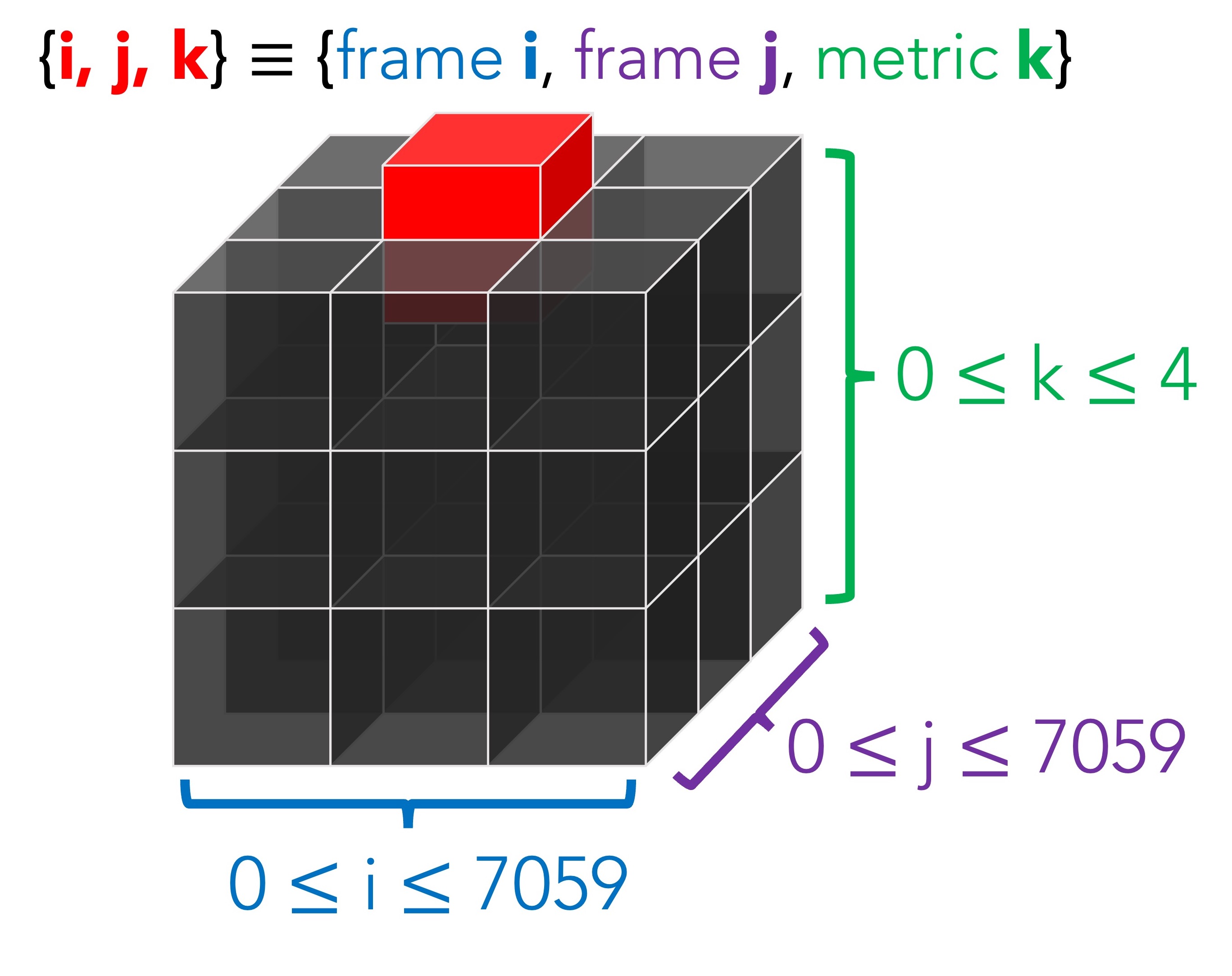}
    \caption{A diagrammatic illustration of the \textit{Metric Covariance Cube}. This represents the data structure used to store image similarity metric values for each pair of science and reference frames. The red cube is an example location in the matrix where the metric value is stored, as described by three indices. The first two indices of the matrix uniquely identify the pair of frames for which the metric is calculated and the third index identifies the metric whose value is stored.}
    \label{fig:covcube}
\end{figure}

\subsubsection{Application to the Sample}
 Identifying the best reference frames based on image similarity metrics during image post-processing requires us to compute and store the metric values for all frames in the sample against every other frame (note that there are no designated reference stars, all reference stars are also science targets in the dataset). The data structure that enables this is a covariance-style three-dimensional matrix of metric values called the \textit{Metric Covariance Cube} (Figure~\ref{fig:covcube}). Since there are 7060 frames in our library and five image similarity metrics, we build a $7060\times7060\times5$ matrix of metric values. The first two indices identify the two frames for which the metric is calculated. The third index identifies which metric is calculated. Building this matrix as a whole is computationally expensive. The computation time can be significantly cut down by recognizing that the matrix is symmetric for any given image similarity metric. The order of the frames does not change the computed metric values. To speed up the remaining computations, we make use of the Caltech high-performance computing cluster (HPC) to run parallelized metric computations. The advantage of building a single matrix of metric values is that during PSF subtraction, it is inexpensive to obtain the metric-ordered set of best matched reference frames for a given target and a given metric. Moreover, it provides the flexibility of either selecting a single metric-ordered reference library for all frames of a given science target star or selecting a different metric-ordered reference library for each science target frame. We note that MSE, PCC, and SSIM are calculated over a frame size of $101 \times 101$ pixels ($\approx$1$\arcsec$ field-of-view across each axis). Since FLSI and CLSI are specifically designed to probe the intensity distribution in coronagraphic images at close separations, they are derived based on mean flux/raw contrast curves calculated over a fixed range of separations from 0$\farcs$05 to 0$\farcs$3, which encompasses the post-coronagraphic PSF pattern.

\subsection{Free Parameter Optimization with Injection-Recovery Tests}
\label{sec:optim}
The application of RDI involves a number of free parameters: (1) image similarity metric by which pre-frame ordering is performed; (2) the number of reference frames (library size) used to construct the principal component (PC) modes; and (3) the number of PCs used to model the stellar PSF for PSF subtraction in the implementation of the Kahrunen Loève Image Processing \citep[KLIP:][]{2012ApJ...755L..28S} algorithm. Computationally, it is impractical to reduce all targets in the sample using every possible combination of image similarity metric, library size, and PCs. To systematically investigate the performance of RDI with varying choice of metrics ($m$), library size ($l$), and number of PCs ($p$), we use synthetic companion injection-recovery tests on a random subset of 50 targets in our sample. The results can provide a good estimate of the optimal set of free parameters that will improve detection sensitivity to point sources for our complete sample. The steps followed are described below.

\begin{figure*}
    \centering
    \includegraphics[scale=0.064]{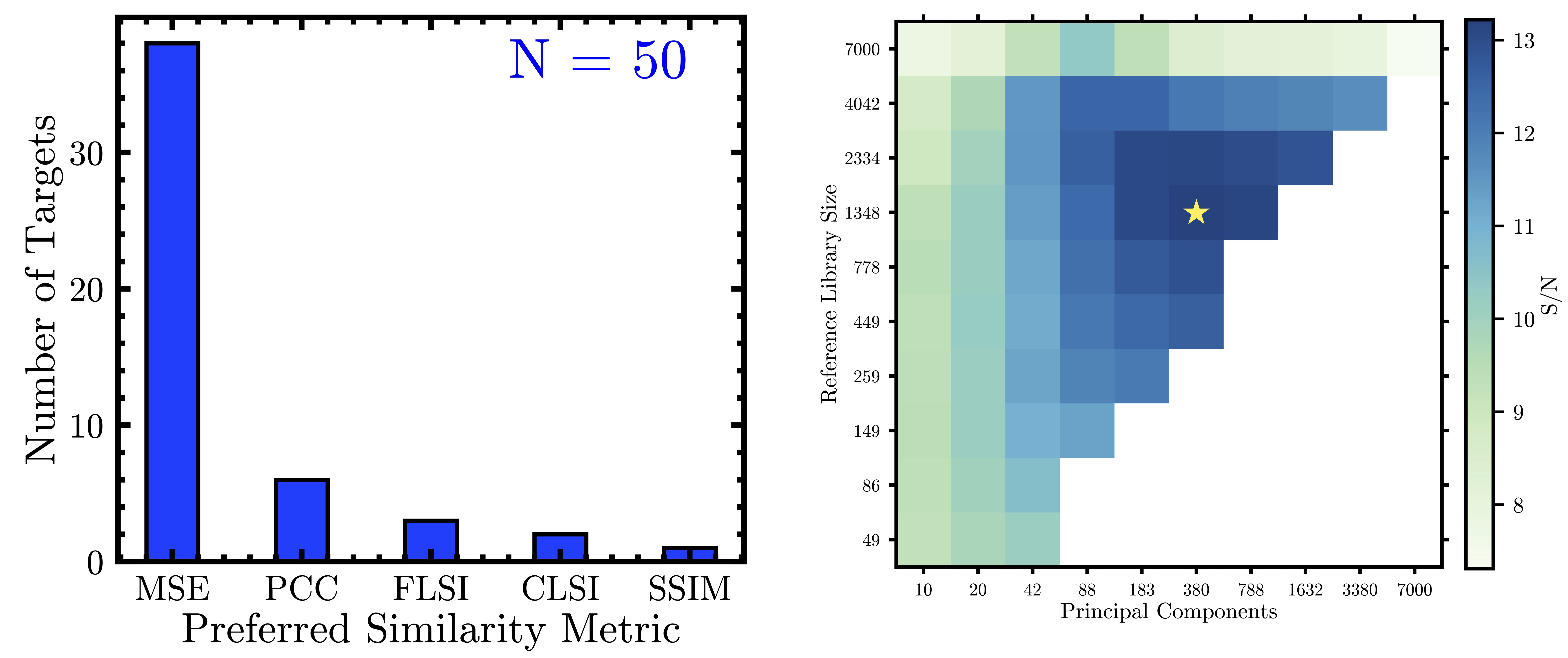}
    \caption{Free parameter optimization results. \emph{Left:} Histogram depicting the number of targets in the injection-recovery subset that yield the highest average S/N for the injected synthetic companion when a given image similarity metric is used for pre-frame ordering. MSE is strongly preferred accounting for random Poisson uncertainty. \emph{Right:} S/N map of the injected synthetic companions as a function of library size and principal components (PCs). Combinations where the number of PCs exceeds the number of reference frames in the library are set to a NaN value. The yellow star indicates the combination of library size and number of PCs that yields the maximum S/N in our tests.} 
    \label{fig:strategy}
\end{figure*}

\begin{enumerate}
    \item For a given science target, we carry out PSF subtraction using a full frame PCA-based approach implemented with the \texttt{pca\_fullfr} function in the Vortex Image Processing \citep[\texttt{VIP};][]{2017AJ....154....7G, Christiaens2023} package. KLIP subtraction is performed using a fiducial set of parameters $(m_0 = \mathrm{PCC},\;l_0 = 500,\;p_0 = 10)$: the frames in the reference library are arranged in descending order of their PCC values, the first 500 frames are selected for PSF modeling, and 10 PCs are used for PSF subtraction.
    
    \item The speckle noise is measured as a function of separation (between 2--5 $\lambda/D$) from the host star in the PSF-subtracted image incorporating small-sample corrections as described in \citet{2014ApJ...792...97M}. Using the speckle noise measurement, we determine the flux of a point source that would yield a 10$\sigma$ detection following similar post-processing, as a function of separation.

    \item A synthetic companion is injected at a given separation and position angle in the science target frames using \texttt{VIP}'s \texttt{cube\_inject\_companions} function. We normalize the unocculted PSF observation for the given science target and use it as a PSF template for injection. We inject point sources for all combinations of four separations (2, 3, 4, and 5 $\lambda/D$) and four position angles ($0^\circ$, $90^\circ$, $180^\circ$, $270^\circ$). This results in a set of 16 point source-injected science target frames. For each combination, we perform PSF subtraction procedures with the fiducial parameters yielding a total of 16 PSF-subtracted images.

    \item The synthetic companion is recovered in each of the PSF-subtracted images. We measure the synthetic companion's flux in each PSF-subtracted image using aperture photometry (1 FWHM diameter aperture). This enables a calculation of the KLIP algorithm's throughput as the ratio of the recovered flux and injected flux (computed in step 2) for each combination of separation and position angle of injection. 

    \item A throughput correction is applied to the flux levels determined in step 2 based on the calculations in step 4. The synthetic companions are then re-injected at the same separations and position angles as before in our science target frames. Steps 1--4 ensure that the synthetic companions are injected at a fixed signal-to-noise ratio (S/N) across all separations and position angles. Thus, when we explore the free parameter space in the next step, any changes in the S/N of the injected synthetic companions can be attributed to changes in our choice of free parameters.

    \item PSF subtraction is performed on all 16 sets of throughput-corrected synthetic companion-injected target frames over the complete free parameter space $(m, l, p)$ (ranges discussed below). The S/N of each recovered synthetic companion is measured to quantify detection significance. This serves as an indicator to the performance of RDI for the tested set of free parameters. 

\end{enumerate}

\noindent
For our dataset, we explored the following free parameter ranges: all five metrics (MSE, PCC, SSIM, FLSI, CLSI), 10 equally logarithmically-spaced library sizes ranging from 50 to 7000 frames, 10 equally logarithmically-spaced principal component values ranging from 10 to 7000 PCs. We cannot generate more principal components than the library size. Thus, library size and principal component pairs where the number of principal components exceeded the library size were skipped and the S/N was set to a NaN value.

Following step 6, we investigate the S/Ns of the retrieved synthetic companions as a function of free parameters $(m, l, p)$ to identify the optimal set of parameter combinations that maximize the detection significance. We conduct this analysis in two steps. First, we determine the best-performing image similarity metric. This metric will be used to reduce the full young M-star survey sample. For a given target, we marginalize across parameters $l$ and $p$ (including separations and position angles of injected sources) by averaging the S/Ns obtained from injection-recovery tests. The metric that then yields the highest average S/N for a given science target is recorded as the preferred similarity metric. This calculation is repeated for all 50 targets in the injection-recovery sample subset and a histogram of metric preferences is generated (left panel; Figure~\ref{fig:strategy}). 38 of the 50 targets prefer MSE for pre-frame ordering. If our dataset did not statistically favor a particular image similarity metric for pre-frame ordering, the number of targets that prefer MSE in the above experiment for a random sample of 50 stars should follow a binomial distribution ($X$) with parameters ($n=50$, $p=0.2$). For this distribution, the corresponding false alarm probability $P(X \ge 38) = 2.5 \times 10^{-17}$ is extremely small. Thus, accounting for random statistics, MSE is strongly preferred by our injection-recovery subset. We can confidently adopt MSE for pre-frame ordering of reference frames for the RDI reduction of the full young M-star survey sample.

Next, given the choice of the MSE metric, we explore the 2D parameter space defined by the reference library size (parameter $l$) and the number of PCs (parameter $p$). For given $(l, p)$, we marginalize across all separations of injection, position angles of injection, and 50 targets by averaging the S/Ns obtained from injection-recovery tests only for the MSE metric. This procedure yields a 2D S/N map as a function of $l$ and $p$ and encapsulates the information from all targets in the subset (right panel; Figure~\ref{fig:strategy}). Examining the S/N map, we find a local S/N maxima at a library size of 1348 frames and 380 PCs. Our results reflect that small library sizes do not sufficiently capture the PSF structure whereas very large library sizes include poor matching reference frames which generate PC modes that add noise to the data. We favor higher number of PCs since they fit more features in the PSF and enable better subtraction results.

These results allow us to define the set of library sizes and PCs to be adopted for the uniform processing of all targets in the sample. For this purpose, we select three library sizes: 1000, 1750, and 3000 frames. These are chosen to sample the localized high S/N region (Figure~\ref{fig:strategy}) while balancing post-processing computational costs: adding new library size options significantly increases the computing time. In contrast, the addition of new PC options does not significantly increase the computing time. Thus, we only adopt an upper limit of 500 PCs based on the S/N map (Figure~\ref{fig:strategy}). Testing a range of PCs is important to understanding the evolution of speckles in the PSF-subtracted images and can enable their differentiation from true astrophysical sources. Additionally, the optimal number of PCs needed for recovery of an astrophysical source can vary with separation of the source \citep[e.g.,][]{2014ApJ...780...17M, 2024arXiv240601809B}.

It is interesting to note here that if we would have restricted ourselves to using references frames from the same night as the target, we would have only been able to compile reference libraries of sizes, at maximum, a few hundred frames. This is sub-optimal based on our injection-recovery tests. Further, we emphasize that our optimized parameters may not generally apply to datasets other than the one presented in this work. The choice of optimal metric, library size, and principal components is highly sensitive to the quality of reference stars obtained and thus may yield different results for different observational samples. Nevertheless, the Super-RDI framework developed in this work can be universally applied to accurately identify the optimal set of reduction parameters for a given dataset. 

\subsection{Uniform Post-Processing and Contrast Curve Determination}
The synthetic companion injection-recovery tests enabled the determination of the optimal set of free parameters for the reduction of the full sample. The adopted parameters were: MSE metric for pre-frame ordering, library sizes $l$ = 1000, 1750, 3000 frames, and $p$ = 1, 5, 10, 20, 50, 100, 200, 300, 400, 500 PCs. Additionally, to efficiently search for close-in and wide-separation companions, we choose three combinations of full frame ($f$) and numerical central mask ($cm$; centered on the host star) sizes for the PCA-based PSF subtraction procedures: (1) $101 \times 101$ pixels frame with a 1 FWHM ($\approx$80 mas) central mask; (2) $201 \times 201$ pixels frame with a 2 FWHM central mask; and (3) $551 \times 551$ pixels frame with a 5 FWHM central mask. For the $551 \times 551$ pixels frame size, we only reduce targets with 1750 reference frames due to the increased computational expense. We uniformly reduce the entire young M-star survey dataset (195 observations of 157 unique targets) for all combinations of $(l, p, f, cm)$ using \texttt{VIP}'s \texttt{pca\_fullfr} function. Simultaneously, after every reduction, we compute the $5\sigma$ contrast curve for the final PSF-subtracted image using \texttt{VIP}'s \texttt{contrast\_curve} function. This function also calibrates the contrast curves for the KLIP algorithm's throughput losses through synthetic companion injection and retrieval. Two data products are generated per science target: 70 PSF-subtracted images and 70 $5\sigma$ contrast curves (3 library sizes $\times$ 10 PCs each for the $101 \times 101$ pixels and $201 \times 201$ pixels reductions and 1 library size $\times$ 10 PCs for the $551 \times 551$ pixels reductions). Each target was reduced on the HPC and took an average of 4 hours to complete on a single node with 32 processes. We note here that 3 out of the 195 science observations in the young M-star survey sample were omitted when performing the above reduction procedures either because of data processing errors in the pipeline or the lack of an unocculted PSF: [SLS2012]~PYC~J00390+1330 (2018~July~30), 2MASS~J06135773-2723550 (2017~January~15), and 2MASS~J14190331+6451463 (2016~March~25). These omissions do not impact subsequent performance analysis. We provide the Super-RDI reduced images for all targets in the young M-star survey on Zenodo\footnote{\url{https://zenodo.org/records/12747613}} for public access. Finally, we create a single ``principal" Super-RDI contrast curve for each target. This is done by combining the contrast curves from all three frame sizes and using the deepest contrast value at each separation across all reductions (where $l$ and $p$ vary). The Super-RDI principal contrast curves cover a separation range of 0$\farcs$1--2$\farcs$63.

\section{Angular Differential Imaging}
\label{sec:adi}
We also reduce the full young M-star survey sample with a widely used implementation of ADI-based PSF subtraction and compute the corresponding $5\sigma$ contrast curves using \texttt{VIP} for comparison with Super-RDI. Several studies in the past have explored optimizing the performance of ADI similar to our exploration for RDI \citep[e.g.,][]{2007ApJ...660..770L, 2023AJ....165...57A}. There are three basic PCA-KLIP parameters that influence the contrast achieved by ADI: (1) the fitting area or search zone over which PCA-KLIP is independently applied; (2) a rotation gap criterion which excludes frames with field rotation less than the specified value to reduce the impact of self-subtraction; and (3) number of principal components (PCs) used to construct the PSF model. A complete exploration of the free parameter space above is outside the scope of this work. For our dataset we use the following combination of free parameter settings.

We adopt similar frame size and central mask settings as our Super-RDI optimizations: $101\times101$ pixels with a 1 FWHM ($\approx$80 mas) mask and $551\times551$ pixels with 5 FWHM mask. We did not consider $201\times201$ pixels reductions with a 2 FWHM mask as they did not improve contrast compared to the above two settings based on tests with a representative sample of our targets. For reductions with a frame size of $101\times101$ pixels and a 1 FWHM central mask, we generate contrast curves for both full frame and annular PCA-KLIP reductions of the sample. For the annular PCA-KLIP contrast curve calculations, we fix the radial size of each annulus to be 1 FWHM and apply a rotation gap criterion of 0 (no frame exclusion), 0.1, 0.5, 1, and 2 FWHM. Due to the low P.A.~rotation of our dataset, we could apply the 0.1, 0.5, 1, and 2 FWHM rotation gap criterion only for 121, 4, 2, and 1 observation (out of 195) respectively in our sample without excluding all available frames. For each of the above settings, we generate contrast curves for a variable number of PCs generally ranging from 1 PC to the number of ADI frames in steps of 1 or 2 PCs. Since our primary goal is to compare the performance of ADI with Super-RDI at small angular separations, we only perform full frame PCA-KLIP reductions for the $551\times551$ pixels frame size and 5 FWHM central mask setting with variable number of PCs. We also note that there is a significant computational expense associated with parameter optimization for this frame size. Appendix~\ref{app} provides more information about the outcomes of parameter optimization with the $101\times101$ pixels reductions in the context of our survey dataset.

We note that there are two improvements that can be made to the above widely used implementation of ADI that may affect sensitivity to point sources. First, the zone geometry in annular PCA-KLIP reductions can be further optimized by dividing the annuli into wedge-like sectors \citep[e.g.,][]{2007ApJ...660..770L}. This may help by providing a less underdetermined fit to the PSF. We did not experiment with this setup since we find that full frame ADI generally performs better than annular ADI for our observations (Appendix~\ref{app}). Second, similar to Super-RDI, it is possible to implement correlation-based frame-selection in ADI post-processing. It has been previously demonstrated that this can boost contrast performance and the S/N of point sources \citep{2010SPIE.7736E..1JM, 2012ApJ...755L..34C, 2014ApJ...795..133C, 2016ApJ...824..117P}.  However, given the low number of ADI frames available per sequence ($\approx$20) in our dataset, excluding additional frames is not likely to provide significant improvements in sensitivity for our observations. We keep as many science frames per target as possible to average temporary alignment errors, AO instability, and sky transmission variations. We only discarded frames with obvious issues (e.g., star not behind the vortex coronagraph).

\begin{figure}
    \centering
    \includegraphics[scale=0.35]{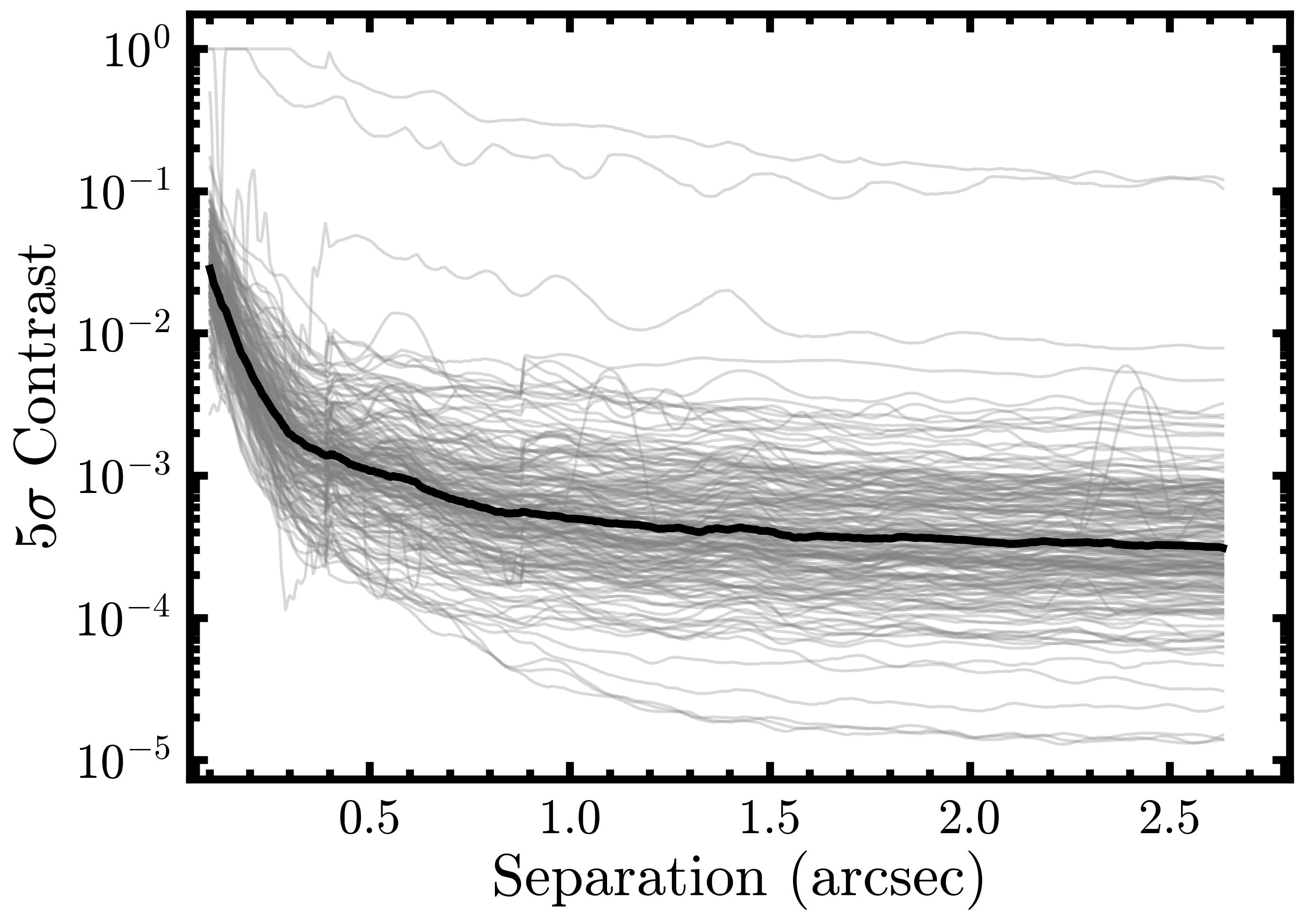}
    \caption{``Optimal" 5$\sigma$ contrast curves for 192 observations in the young M-star survey (gray curves) constructed by selecting the deepest achieved contrast as a function of separation from the combination of Super-RDI and ADI principal contrast curves. The median optimal contrast curve is plotted in black. A few contrast curves show irregularities due to the presence of point sources or image artifacts.}
    \label{fig:cc}
\end{figure}

We create a single ``principal" ADI contrast curve for each target. This is done by combining the contrast curves from the two frame sizes and using the deepest contrast value at each separation across all reductions (where the fitting area, rotation gap criterion, and number of PCs vary as described above). The ADI principal contrast curves cover a separation range of 0$\farcs$1--2$\farcs$63. Finally, we also combine the Super-RDI and ADI principal contrast curves for each target by selecting the deepest contrast value at each separation to generate a final ``optimal" contrast curve. These are plotted in Figure~\ref{fig:cc} along with the median optimal contrast curve computed across all targets. A few contrast curves show irregularities due to the presence of point sources or image artifacts. These comprise a negligible percentage of the total sample and do not impact our analyses. The final optimal contrast curves can be used to incorporate our results in future substellar demographic studies. We provide these on Zenodo\footnote{\url{https://zenodo.org/records/12747613}} for public access. Optimal contrasts at select separations are included in Table~\ref{tabA1} in Appendix~\ref{app:B}.

\section{Super-RDI Performance Characterization}
\label{sec:char}
To analyze the performance of Super-RDI, we use the $5\sigma$ contrast curves computed previously as our primary tool. They are a measure of our sensitivity to companions after PSF subtraction. In this section, we (1) compare the contrast achieved by Super-RDI against ADI; (2) investigate the dependence of the performance of Super-RDI in comparison to ADI on the P.A.~rotation of our dataset; and (3) examine Super-RDI's sensitivity as a function of host star \emph{W1} and \emph{R} magnitude\footnote{We note here that analysis of contrast as a function of seeing amplitude did not yield any correlations as has been previously found in literature \citep{2016SPIE.9909E..0VB, 2018AJ....156..156X}.}.

\subsection{Super-RDI vs ADI for the Young M-star Survey}
\label{sec:comparison}

\subsubsection{Preamble}
Here, we use the $5\sigma$ Super-RDI and ADI principal contrast curves to determine which strategy performs better as a function of separation from the host star for observations obtained in the young M-star survey. There are certain aspects of our dataset that must be considered before interpreting the results in this section.

\begin{enumerate}

    \item The young M-star survey observations have a median P.A.~rotation of 8.6$^\circ$ and a mean P.A.~rotation of 16.4$^\circ$. This corresponds to a physical rotation of 0.38 FWHM and 0.73 FWHM at $0\farcs2$ separation and 0.76 FWHM and 1.46 FWHM at $0\farcs4$ separation respectively. Thus, ADI reductions are expected to suffer from self-subtraction effects for a significant number of targets in the sample at small angular separations. The performance comparison of Super-RDI with ADI as a function of P.A.~is discussed in detail in the next section.
    
    \item The typical number of exposures for targets in the sample is $\sim$20. For a 1 FWHM width annulus at $0\farcs2$ and $0\farcs4$ separation, there are $\approx$15 and $\approx$30 independent speckle realizations (where a speckle typically has size $\sim$1 FWHM), respectively. Thus, there are cases where PSF fitting with observations in our ADI sequences may be underdetermined even at small angular separations. In contrast, a 101$\times$101 pixels full frame size includes $\sim$200 independent speckle realizations and our Super-RDI reductions use reference library sizes that are typically much larger (1000--3000 frames) for PSF fitting.
    
\end{enumerate}

These characteristics of the dataset lead us to expect that Super-RDI will perform better than ADI in the speckle-dominated noise regime (small angular separations) for our observations. As discussed in \S\ref{sec:obs-strat}, the above characteristics of our dataset are a result of the design of the young M-star survey under the practical considerations of observing a large number of targets with limited availability of telescope time (which determines the number of exposures obtained and P.A.~rotation achieved) and scheduling conflicts within a given night (which determines if targets can be observed near transit, where the P.A.~rotation is largest). 

In this context, the goal of the following analysis is to understand where the performance transition between ADI and Super-RDI occurs for young M-star survey observations in separation space. These results can inform the design of future high-contrast imaging surveys. Such analyses have previously been conducted in literature. For example, \citet{2017AJ....154...73R} provided the first perfomance comparison between RDI and ADI with the Keck/NIRC2 vortex coronagraph in imaging observations of a single target, TW~Hya, where they achieved $\approx$45$^\circ$ rotation on three separate nights. They found that RDI (using designated reference stars imaged before and after the science target) achieves better contrast than ADI at separations $\lesssim 0\farcs25$ when data from all three nights were combined. \citet{2018AJ....156..156X} used a significantly larger Keck/NIRC2 $L'$ and $M_s$ dataset consisting of 359 observations of 304 unique stars, with median and mean P.A. rotations of 11.1$^\circ$ and 26.0$^\circ$ respectively, and found that RDI performed better than ADI at separations $\lesssim 0\farcs25$, averaged across all the targets in their sample. In their work, RDI was performed using designated reference stars or all stars observed on the same night as the science target. We note here that the observations used by \citet{2018AJ....156..156X} in their analyses include all the observations presented in this work as part of the young M-star survey. This allows us to make a more direct comparison between Super-RDI and the implementation of RDI using same-night reference star observations. 

\begin{figure*}
    \centering
    \includegraphics[scale=0.16]{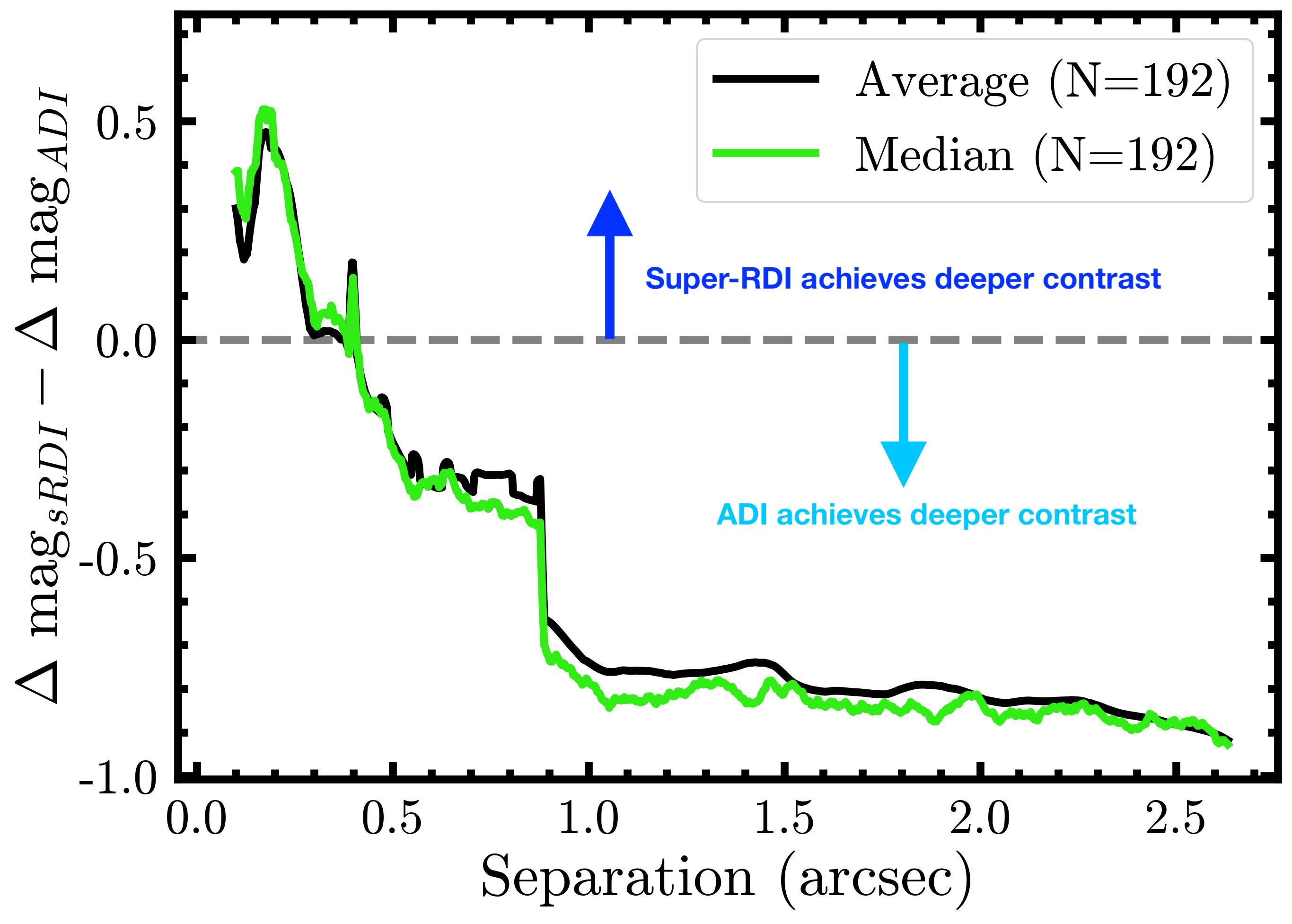}
    \caption{\emph{Super-RDI vs ADI for the young M-star survey:} Average (black) and median (green) difference in Super-RDI (sRDI) and ADI contrast (units of $\Delta$mag) as a function of separation (units of arcseconds). A gray dashed line marks zero difference and corresponds to both techniques achieving the same sensitivity level. Positive differences indicate better Super-RDI performance and negative differences indicate better ADI performance. For the typical parallactic angle rotation of the young M-star survey ($\sim$10$^\circ$), Super-RDI outperforms a widely used implementation of ADI-based PSF subtraction at separations $\lesssim 0\farcs4$, averaged across all observations.}
    \label{fig:comp}
\end{figure*}

\subsubsection{Contrast Curve Comparison}
First, contrast is converted to units of $\Delta$mag. Next, we compute the average and median of the $5\sigma$ Super-RDI and ADI principal contrast curves across all targets. Taking the difference of the corresponding Super-RDI and ADI average and median principal contrast curves provides an estimate of the difference in sensitivity achieved between the two techniques as a function of separation. We plot this difference in Figure~\ref{fig:comp}. For the young M-star survey, Super-RDI outperforms a widely used implementation of ADI-based PSF subtraction at separations $\lesssim 0\farcs4$ ($\approx$5$\lambda$/D), gaining as much as 0.25\,mag in contrast at $ 0\farcs25$ and 0.4\,mag in contrast at $0\farcs15$. The contrast gain achieved by Super-RDI increases as the separation decreases. ADI outperforms Super-RDI for separations $\gtrsim 0\farcs4$. The location of the transition separation in contrast performance is determined by the rate of ADI's decline in performance due to self-subtraction effects in comparison to the rate of RDI's decline in performance due to over-subtraction effects. Compared to the implementation of RDI with same-night reference star observations in \citet{2018AJ....156..156X}, the Super-RDI framework represents a performance improvement in separation space, from previously achieving deeper contrasts compared to ADI at separations $\lesssim 0\farcs25$ to now achieving deeper contrasts compared to ADI at separations $\lesssim 0\farcs4$. Super-RDI reduces speckle noise in the PSF-subtracted images through metric-based frame selection and finds the optimal number of KLIP principal components by maximizing the S/N of synthetic companions in our injection-recovery tests. This helps reduce the impact of over-subtraction and yields improvements in RDI's performance at small angular separations. In Figure~\ref{fig:comp}, we observe a step-like feature at $\approx$$0\farcs9$. This is a consequence of the frame sizes selected in our reduction pipeline. The step-like feature occurs at the contrast calculation boundary of the $201 \times 201$ pixels frame reductions. It indicates a transition to contrast values obtained with the $551 \times 551$ pixels frame reductions, where the performance of Super-RDI degrades significantly. This is likely because the similarity between reference and science frames is poorer at wide angular separations from the central speckle halo (the region over which the metrics are computed). Excluding the step-like artifact discussed above, we observe that the difference in contrasts achieved by the two techniques remains roughly constant beyond $0\farcs6$. This marks the transition from the speckle-dominated noise regime to the background-limited noise regime. The above results demonstrate that Super-RDI is the preferred strategy for the detection of companions at small angular separations ($\lesssim 0\farcs4$) in the context of our survey design, for a dataset with median P.A. rotation of $8.6^\circ$. ADI remains the preferred strategy for companion searches at wider angular separations. 

\begin{figure*}
    \centering
    \includegraphics[scale=0.1]{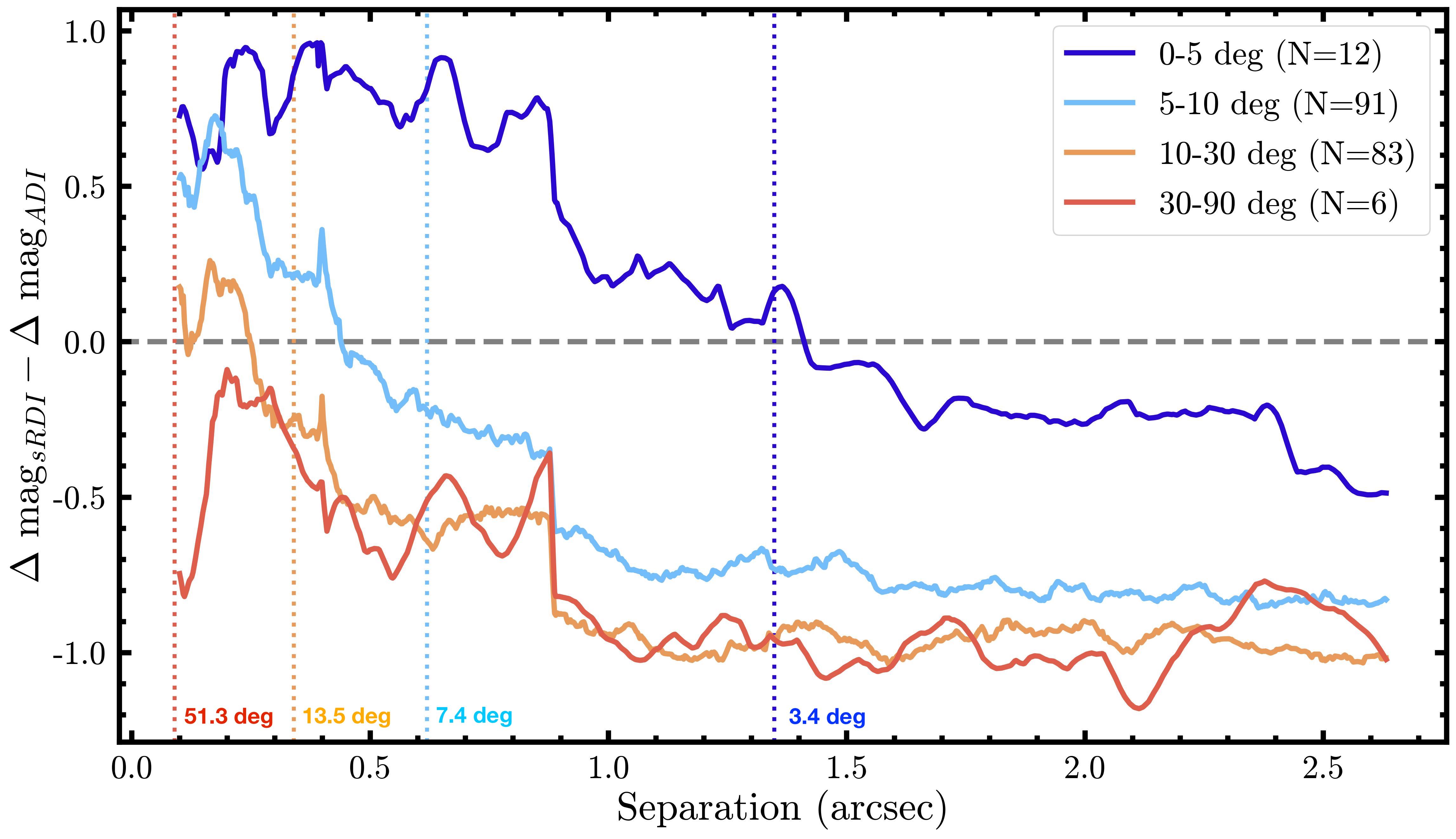}
    \caption{Median difference in Super-RDI (sRDI) and ADI contrast (units of $\Delta$mag) as a function of separation (units of arcseconds) for observations sorted into four $\Delta$P.A. bins (units of degrees). A horizontal gray dashed line marks zero difference and corresponds to both techniques achieving the same sensitivity level. Positive differences indicate better Super-RDI performance and negative differences indicate better ADI performance. Vertical dashed lines mark the separations at which 1 FWHM rotation is achieved for the median $\Delta$P.A. value of targets in each bin. The step-like feature at $\approx$$0\farcs9$ is discussed in \S\ref{sec:comparison}.}
    \label{fig:pa}
\end{figure*}

As we discussed in \S\ref{sec:optim}, it is important to note that the optimal reduction parameters (which directly impact Super-RDI's performance) are specific to the observational sample used in this work. They may change for different datasets. Further, the analysis above is conducted for a group of targets. The performance of Super-RDI may vary between individual targets. In practice, we recommend complementing the application of Super-RDI with dedicated reference star observations to ensure the best performance for science targets.  

\subsection{Dependence on Parallactic Angle Rotation}
\label{sec:pa}
We investigate the dependence of the performance of Super-RDI in comparison to ADI on the P.A.~rotation ($\Delta$P.A.) of our dataset. The targets are sorted into four $\Delta$P.A. bins: 0--5$^\circ$, 5--10$^\circ$, 10--30$^\circ$, and 30--90$^\circ$ (for the young M-star survey dataset, the maximum $\Delta$P.A. $= 89.2^\circ$). Similar to \S\ref{sec:comparison}, we compute the median Super-RDI and ADI principal contrast curve across all targets in a given $\Delta$P.A. bin and present the difference in contrasts between the two in Figure~\ref{fig:pa}. The analysis in \S\ref{sec:comparison} showed that for our dataset, Super-RDI improves upon ADI at separations $\lesssim 0\farcs4$ (averaged across the entire sample). Here, we find that for targets with $\Delta$P.A.~= 0--5$^\circ$, Super-RDI outperforms ADI up to a much larger separation of $1\farcs4$ due to the severe self-subtraction effects experienced by ADI. For targets with $\Delta$P.A.~= 5--10$^\circ$, Super-RDI outperforms ADI at separations $\lesssim 0\farcs4$, consistent with the average in \S\ref{sec:comparison}. For targets with $\Delta$P.A. = 10--30$^\circ$, Super-RDI's advantage in separation space reduces to $\lesssim 0\farcs25$. ADI achieves deeper contrasts across all separations for targets with $\Delta$P.A.~= 30--90$^\circ$. Note that the results for the smallest and largest $\Delta$P.A. value bins are based on a comparatively small sample of objects and should be interpreted with caution. We also observe visual consistency between the separation at which ADI begins outperforming Super-RDI and the separation corresponding to a 1 FWHM rotation for the median $\Delta$P.A. value of targets in each bin in Figure~\ref{fig:pa}. 

Our results indicate that ADI is better suited for companion detection at small angular separations for studies focusing on a limited number of systems where greater scheduling flexibility can allow larger P.A.~rotations to be accrued for similar integration times. However, for large surveys, such as the young M-star survey, the necessity to balance the number of targets observed with limited telescope time availability and scheduling constraints may require an RDI-focused design. In such cases, the Super-RDI framework can help improve the sensitivity of RDI at small angular separations by leveraging a large reference PSF library. 

\begin{figure*}
    \centering
    \includegraphics[scale=0.075]{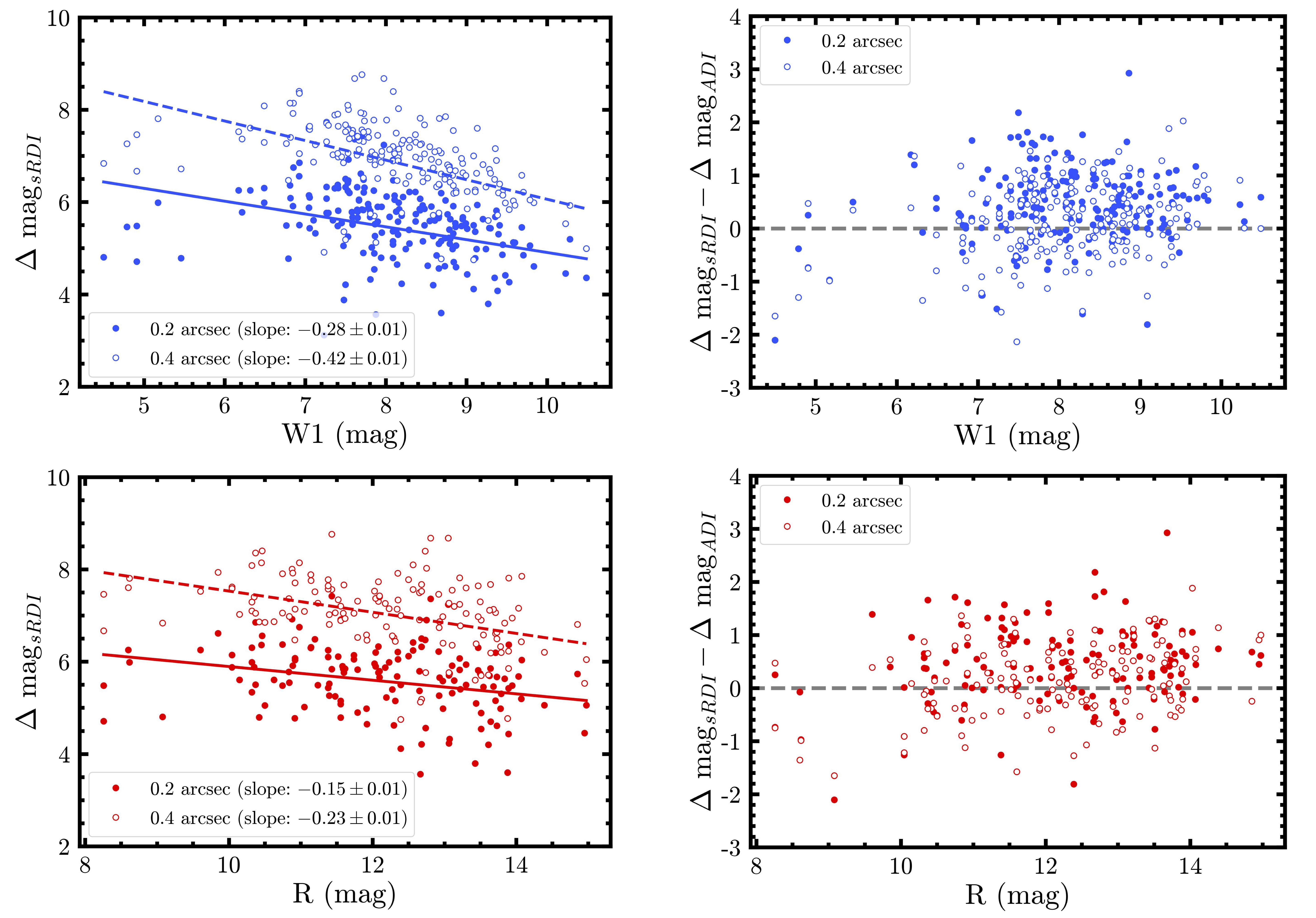}
    \caption{\emph{Left Column:} Contrast (in $\Delta$mag units) at $0\farcs2$ (solid blue/red circle) and $0\farcs4$ (open blue/red circle) plotted as a function of host star \emph{W1}/\emph{R} magnitude for all targets in the sample. The corresponding best-fit lines are plotted as blue/red solid and dashed lines. Negative slopes of the best-fit line indicate decreasing sensitivity for fainter stars. See \S\ref{sec:flux} for discussion on discrepancies of the data with the best-fit line for the brighter targets. \emph{Right Column:} Difference in contrast (in $\Delta$mag units) at $0\farcs2$ (solid blue/red circle) and $0\farcs4$ (open blue/red circle) achieved by Super-RDI (sRDI) and ADI as a function of host star \emph{W1}/\emph{R} magnitude for all targets in the sample. A gray dashed line marks zero difference corresponding to both techniques achieving the same sensitivity level. Super-RDI shows an improvement over ADI for the majority of fainter targets at both separations while the reverse is true for brighter targets.}
    \label{fig:mag}
\end{figure*}

\subsection{Dependence on Stellar Fluxes}
\label{sec:flux}
We investigate the performance of Super-RDI as a function of host star flux using the calculated $5\sigma$ principal contrast curves. We focus on two bandpasses: (1) the host star's \emph{W1} magnitude ($\lambda_{\mathrm{eff}} = 3.35\;\mu\mathrm{m}$, FWHM $= 0.64\;\mu\mathrm{m}$) as a proxy for the $L'$ band ($\lambda_{\mathrm{eff}} = 3.776\;\mu\mathrm{m}$, FWHM $= 0.70\;\mu\mathrm{m}$) in which observations were made; and (2) the host star's \emph{R} magnitude ($\lambda_{\mathrm{eff}} = 0.64\;\mu\mathrm{m}$, FWHM $= 0.16\;\mu\mathrm{m}$) due to the Keck II AO system wavefront sensor's sensitivity to those wavelengths. 

First, contrast achieved by Super-RDI is plotted as a function of the corresponding host star's \emph{W1} and \emph{R} magnitude for each target in the sample. We choose to plot the contrast achieved at two separations within the range where Super-RDI outperforms ADI (averaged across all targets) for our dataset: $0\farcs2$ and $0\farcs4$ (left panels; Figure~\ref{fig:mag}). Visual inspection reveals a trend in the data for both bandpasses and separations which we quantify mathematically by computing slopes using a least squares linear fit (\texttt{scipy.optimize.curve\_fit}). The uncertainty in the fit is the square root of the variance estimate for the slope parameter. We find negative slopes for all best-fit lines to the data (left panels; Figure~\ref{fig:mag}) indicating that Super-RDI's sensitivity decreases as the host star becomes fainter. This is expected since fainter targets are more susceptible to deviations from the ideal instrument PSF structure due to variability in observing conditions and poorer AO performance. Our results are in agreement with a similar analysis conducted in \citet{2018AJ....156..156X}. A closer look at the fits reveal disagreement between the best-fit line and contrast measurements for targets with \emph{W1} magnitudes $<$6\,mag and \emph{R} magnitudes $<$9.5\,mag: Super-RDI achieves shallower contrasts than expected from the best-fit line but comparable contrasts to targets with host star magnitudes just greater than the above values. This may indicate that for targets brighter than these host star magnitudes, the Super-RDI's performance plateaus and does not yield additional gains.

Next, the difference in contrast achieved by Super-RDI and ADI is plotted as a function of the corresponding host star's \emph{W1} and \emph{R} magnitude for each target in the sample (right panels; Figure~\ref{fig:mag}). Similar to above, the contrast values at $0\farcs2$ and $0\farcs4$ are chosen. Such an analysis enables us to determine if at small angular separations there is a preferred strategy depending on the host star's magnitude. We find that for targets with \emph{W1} magnitudes $>$6\,mag and \emph{R} magnitudes $>$9.5\,mag in our survey, Super-RDI outperforms ADI in most cases at both separations. Note that the gain for individual targets can be higher than the sample-averaged values presented in \S\ref{sec:comparison}. Below these magnitudes, ADI outperforms Super-RDI in most cases at both separations. This indicates that ADI should be the preferred strategy for bright targets (magnitude ranges above). Super-RDI is the recommended strategy for fainter targets. Note that the data is consistent with zero slope and thus Super-RDI does not offer increasing or decreasing contrast gains as a function of host star magnitude compared to ADI. 

\section{Point Source Detections}
\label{sec:det}
We search for point sources in our uniformly reduced images of the complete young M-star survey dataset by visual inspection. To avoid mistaking speckles for companion detections, we consider the evolution of candidate point sources as the number of principal components change in the image. A speckle is generally variable in nature as the number of PCs changes whereas a true astrophysical source remains roughly stable in the image. For visually flagged sources, we also compute the S/N to quantify detection significance. The S/N is calculated accounting for small-sample statistics as detailed in \citet{2014ApJ...792...97M}. 

Visual inspection reveals four point source detections from a total of 157 unique targets. We detect point sources around 2MASS~J01225093-2439505, 2MASS~J23513366+3127229, 2MASS~J06022455-1634494, and LO~Peg. Two of the detections are well-characterized substellar companions: 2MASS~J01225093-2439505~B \citep{2013ApJ...774...55B} and 2MASS~J23513366+3127229~B \citep{2012ApJ...753..142B}. As we will discuss in an upcoming section, the candidate companion around 2MASS~J06022455-1634494 is likely a background star based on a common proper motion test. Analysis of the pre-processed coronagraphic images shows that the point source around LO~Peg is a speckle. Forward modeling procedures and point source characteristics are described below.

\subsection{Forward Modeling}
\label{sec:fm}
We perform photometric and astrometric measurements of the point source detections using negative synthetic companion injection. This technique calibrates measurement biases introduced in the primary subtraction procedures. We apply this method to our observations using \texttt{VIP}'s \texttt{firstguess} function. First, synthetic companions (normalized unocculted PSF of the target) are injected in a grid near the location of the detection at varying \emph{negative} flux values. The reduced $\chi^2$ value, calculated after PSF subtraction of the negative synthetic companion-injected frames, is minimized at the location of injection to obtain a first estimate of the astrometry and photometry of the point source. The estimated parameters are used as inputs to a simplex Nelder-Mead minimization algorithm \citep[][implemented with argument \texttt{simplex=True}]{nelder_simplex_1965} which provides a more accurate estimate of the point source's separation, position angle, and flux. We use these values as our forward modeled parameters. We provide \texttt{firstguess} with initial position coordinates for the grid-based minimization step based on visual inspection of the PSF-subtracted image in SAO DS9\footnote{\url{https://sites.google.com/cfa.harvard.edu/saoimageds9}}, choose an annulus width equal to 1.5 FWHM for PCA subtraction, and define the aperture size for flux calculations to be 1 FWHM.

\subsection{Uncertainty Analysis}
To estimate the uncertainty in our forward modeled astrometry and photometry, we use synthetic companion injection-recovery tests. First, the point source detection is removed from the science image cubes by injecting a negative-flux (same as forward modeled value) synthetic companion at the forward modeled separation and position angle. This is done to prevent contamination by the point source detection in subsequent analyses. Next, synthetic companions are injected at the point source detection's forward modeled radial separation at eight equally spaced position angles. We forward model each injected synthetic companion following the simplex Nelder-Mead minimization method discussed in \S\ref{sec:fm}. Finally, the error in position angle and separation, and the percentage flux change are recorded for each injected companion. To obtain the total astrometric uncertainty, the standard deviation of the errors are combined with the uncertainties in the distortion solution, centering accuracy of the QACITS algorithm \citep{2017A&A...600A..46H}, north alignment, and plate scale following \citet{2022AJ....163...50F}. To estimate the photometric uncertainty, we combine the speckle noise with the KLIP throughput uncertainty. Speckle noise is determined from the standard deviation of the flux integrated within apertures defined in the S/N calculation. The KLIP throughput uncertainty is obtained using the standard deviation of the percentage change in injected synthetic companion flux when injection-recovery tests are performed at the separation of the point source detection at varying position angles.

\input{tab2.tex}

\subsection{Mass Estimates}
We can combine the forward modeled photometry with an age estimate for the host star to obtain the point source's approximate mass (assuming it is a companion) using evolutionary models. Here, we use the COND hot-start evolutionary model grid \citep{2003A&A...402..701B} for our calculations. First, we convert the forward modeled flux estimate ($f_0$; optimization is conducted over a 1 FWHM region) of the point source to a contrast value ($C$) as follows, 
\begin{equation}
    C = -2.5\cdot \mathrm{log}_{10} \left( \frac{f_0}{f_{\mathrm{{FWHM, *}}}} \cdot \frac{T_{\mathrm{int, *}} \cdot N_{\mathrm{*}}}{T_{\mathrm{int, sci}} \cdot N_{\mathrm{sci}}} \right),
\end{equation}
\noindent
where $f_{\mathrm{{FWHM, *}}}$ is the aperture-integrated flux of the host star in a 1 FWHM region, $T_{\mathrm{int, *}}$ is the integration time per coadd, and $N_{\mathrm{*}}$ is the number of coadds, all determined from the host star's unocculted PSF observation. $T_{\mathrm{int, sci}}$ is the integration time per coadd and $N_{\mathrm{sci}}$ is the number of coadds, both determined from the science observations. Next, we estimate the absolute $L'$ magnitude of the host star ($M_{L', *}$) using its apparent \emph{W1} magnitude and a parallax measurement. The absolute $L'$ magnitude of the point source (assuming it is at the distance of the host star) is given by
\begin{equation}
    M_{L', \mathrm{PS}} = M_{L', *} + C.
\end{equation}
\noindent
Next, we create a 2D linear interpolation function for the COND model grid that maps mass and age inputs to absolute $L'$ magnitude outputs using \texttt{scipy.interpolate.interp2d}. We draw an age from the age distribution of the host star and define a set of $10^4$ test mass values ranging from 0.0005 $M_\odot$ to 0.1 $M_\odot$. The COND grid is interpolated using the previously calculated function at each test mass for the drawn age to get an absolute $L'$ magnitude ($M_{L', \mathrm{COND}}$). The test mass value at which the squared difference between $M_{L', \mathrm{COND}}$ and $M_{L', \mathrm{PS}}$ is minimized is our estimate for the point source's mass. This complete procedure is repeated for $10^4$ different age draws. The mean and standard deviation of the resulting mass estimates provides a range of the possible masses for the point source.

\begin{figure*}
    \centering
    \includegraphics[scale=0.030]{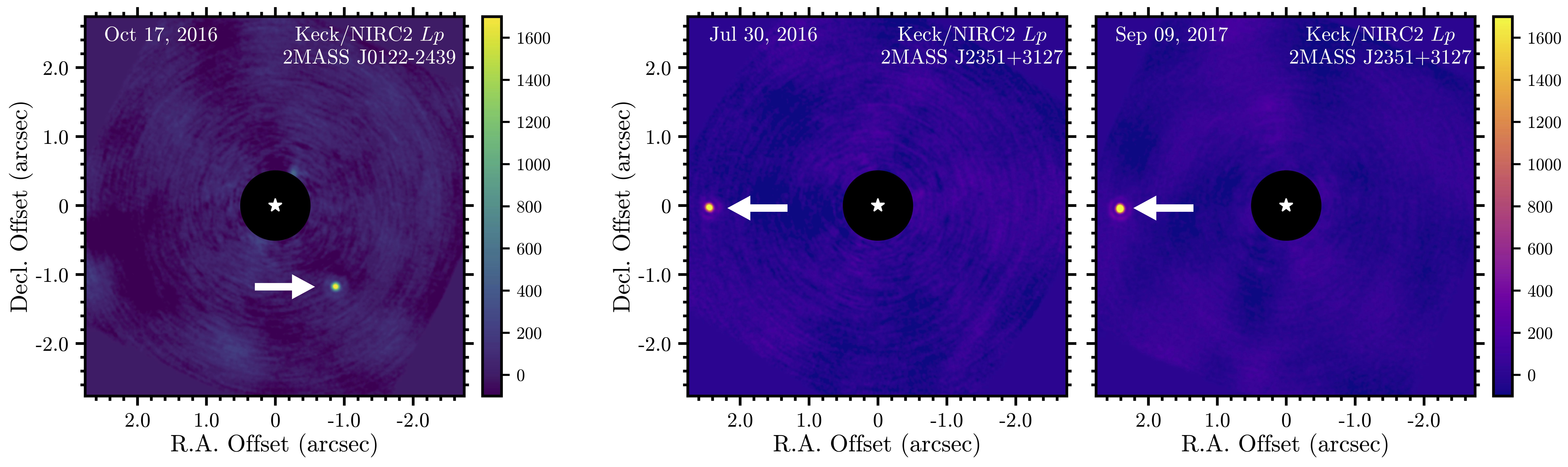}
    \caption{Previously known companions detected with Super-RDI in Keck/NIRC2 $L'$ observations. The final PSF-subtracted images of 2MASS J0122-2439 (one epoch; left) and 2MASS J2351+3127 (two epochs; right) are presented above. White arrows point to the detected companions. North is up and east is to the left. All three images and the color bars are depicted in linear scale in analog to digital units and have been smoothed using a Gaussian kernel with a 1.5 pixel standard deviation (40\% of the width of the instrumental PSF) to average out high-frequency noise.}
    \label{fig:detections}
\end{figure*}

\input{tab3.tex}

\subsection{Known Companions}
\subsubsection{2MASS J01225093-2439505}
\label{2M0122}
2MASS~J01225093-2439505 is an active M3.5V star \citep{2013ApJ...774...55B} with high X-ray luminosity and strong H$\alpha$ emission \citep[][EW$_{\mathrm{H}\alpha}$ = 9.7 \AA]{2006AJ....132..866R}. It is a likely AB Doradus young moving group \citep[$149_{-19}^{+51}$ Myr;][]{2015MNRAS.454..593B} member based on a BANYAN-$\Sigma$ probability of 87.9\% \citep{2018ApJ...856...23G} computed using its parallax, proper motion, and radial velocity (Table~\ref{tab2}). \citet{2013ApJ...774...55B} discovered an L-type companion 2MASS~J01225093-2439505~B ($M_{L'} = 10.4 \pm 0.3$\,mag) at a projected separation of $1\farcs452 \pm 0\farcs005$ and position angle (PA) of $216.6^\circ \pm 0.4^\circ$ ($L'$ observations with NIRC2 on 2013~January~19). This companion falls in the deuterium burning region of the hot-start evolutionary model grids where tracks overlap resulting in a dual-valued mass prediction of 12--13 $M_{\mathrm{Jup}}$ or 22--27 $M_{\mathrm{Jup}}$ based on the host star's AB Dor membership. 

We observed 2MASS~J01225093-2439505 on 2016~October~17~UT and detected the substellar companion (Figure~\ref{fig:detections}) at a S/N of 38. Our astrometric and photometric results for the companion as well as its mass estimate are presented in Table~\ref{tab3}. The photometry and mass are consistent with those of \citet{2013ApJ...774...55B}. We find that the mean separation and PA~are discrepant with orbital fits to the compilation of literature astrometry for the companion presented in \citet{2020AJ....159..181B} by $\approx$8 mas and $\approx$1.5$^\circ$ respectively. The magnitude of the discrepancy is consistent with a possible error in star centering behind the vortex coronagraph. This is particularly important to consider for short sequences (such as this observation: 15 min on-source integration time) where positional offsets in QACITS may not average out as in longer sequences. 

To estimate the drift in star center with respect the vortex center, we can use the fact that the companion is expected to remain at a fixed position (with random noise) in the individual \emph{de-rotated} frames. We fit a 2D Gaussian to the companion PSF in each of the sky-subtracted de-rotated image frames to estimate the source centroid. We find there is an unusually high systematic drift of $\approx$80 mas along the image x-axis and $\approx$20 mas along the image y-axis in the position of the companion across the observation sequence. It should be noted that the pointing accuracy and stability provided by QACITS is 4.5 mas on average and 2.4 mas rms respectively \citep[for details, refer to][]{2017A&A...600A..46H}. The observed drift is thus atypical for vortex observations at Keck and could be due to a combination of relatively poorer seeing ($\approx$0.9) on the night and the target being faint ($R$ = 13.890\,mag). 

We make a conservative estimate of the systematic error in separation and PA by drawing x and y offsets of the star behind the vortex coronagraph from normal distributions with width equal to the estimated drift and adding them to our forward modeled values. The standard deviation of the resulting separation and PA distributions provide estimates of the systematic errors. These are added in quadrature with the measurement uncertainty and reported in Table~\ref{tab3}.

\begin{figure*}
    \centering
    \includegraphics[scale=0.06]{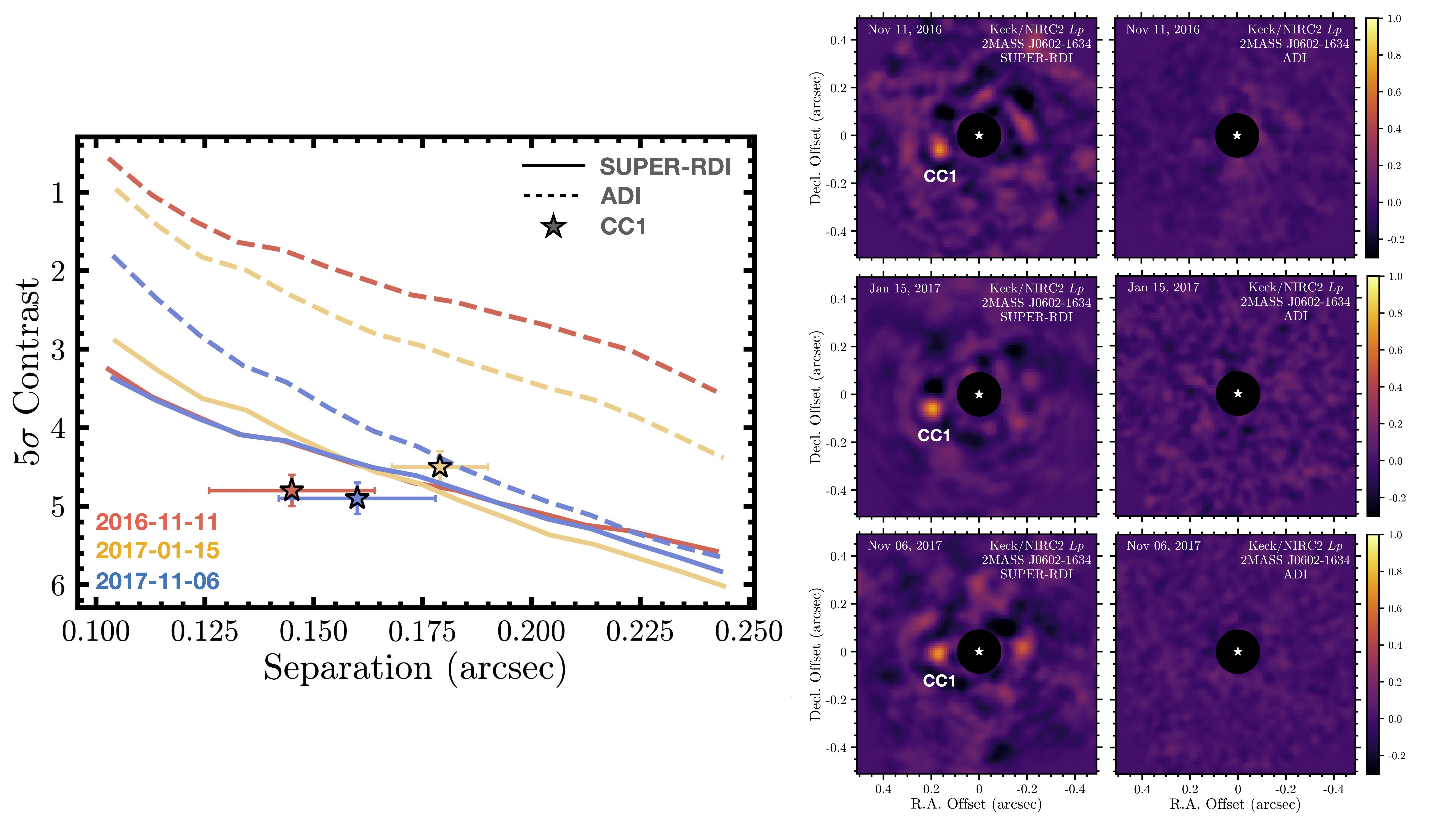}
    \caption{\emph{Left:} 5$\sigma$ Super-RDI and ADI contrast curves for 2M0602-1634 for each observation epoch. The point source detection (``CC1") is marked with a star. \emph{Right:} Final PSF-subtracted images of 2M0602-1634 obtained with Super-RDI and ADI for each observation epoch. North is up and east is to the left. The two images and the color bars are depicted in linear scale in normalized analog to digital units and have been smoothed using a Gaussian kernel with a 1.5 pixel standard deviation (40\% of the width of the instrumental PSF) to average out high-frequency noise.}
    \label{fig:06022}
\end{figure*}

\subsubsection{2MASS J23513366+3127229}
2MASS~J23513366+3127229 is a young M2.0V star with a large X-ray flux \citep{2006AJ....132..866R, 2009ApJ...699..649S}. It is an AB Doradus young moving group \citep[$149_{-19}^{+51}$ Myr;][]{2015MNRAS.454..593B} member based on a BANYAN-$\Sigma$ probability of 99.5\% \citep{2018ApJ...856...23G} computed using its parallax, proper motion, and radial velocity (Table~\ref{tab2}). \citet{2012ApJ...753..142B} discovered an $\sim$L0 substellar companion 2MASS~23513366+3127229~B at a projected separation of $2386.3 \pm 1.5$ mas and a position angle of $91.81^\circ \pm 0.04^\circ$ (\emph{K'} observations with NIRC2 on 2011.871). The companion's mass is $32 \pm 6\;M_{\mathrm{Jup}}$ \citep{2012ApJ...753..142B}. 

We observed 2MASS~J23513366+3127229 on 2017~September~9~UT and 2018~July~30~UT and detected the substellar companion (Figure~\ref{fig:detections}) at a S/N of 39 and 47 respectively. Our astrometric and photometric results for the companion as well as its mass estimate are presented in Table~\ref{tab3}. The mass is consistent with that presented in \citet{2012ApJ...753..142B}. We find that the astrometry of the companion in the 2018 observation epoch is discrepant in separation by $\approx$30 mas with the trend predicted by the compilation of literature astrometry for the companion in \citet{2020AJ....159...63B}. Following an identical process to that described in \S\ref{2M0122}, we find a systematic drift of $\approx$50 mas ($\approx$24 mas) along the image x-axis and $\approx$34 mas ($\approx$39 mas) along the image y-axis in the position of the companion across the 2018 (2017) observation sequence. This could be due to the target being faint ($R$ = 13.105\,mag). The associated systematic errors are added in quadrature with the measurement uncertainty and reported in Table~\ref{tab3}.

\subsection{New Point Source Detection: 2MASS~J06022455-1634494}
2MASS~J06022455-1634494 is an active M0 star \citep{2006AJ....132..866R} with detections of optical and X-ray flares \citep{2003A&A...403..247F, 2019ApJ...881....9H}. It is a field object with a BANYAN-$\Sigma$ probability of 99.9\% \citep{2018ApJ...856...23G} computed using its parallax, proper motion, and radial velocity (Table~\ref{tab2}). We observed 2MASS~J06022455-1634494 at three difference epochs: 2016.86, 2017.04, and 2017.85. We detect a point source (``CC1") at ($145 \pm 19$ mas, $108.3^\circ \pm 6.1^\circ$), ($179 \pm 11$ mas, $103.8^\circ \pm 4.4^\circ$), and ($160 \pm 18$ mas, $91.3^\circ \pm 3.9^\circ$) at a S/N of $\approx$7 (4$\sigma$ detection), $\approx$8 (5$\sigma$ detection), and $\approx$6 (3.9$\sigma$ detection)\footnote{The S/N ratio, computed following \citet{2014ApJ...792...97M}, is converted to a Gaussian detection significance level for the equivalent false positive probability using the \texttt{significance()} function in \texttt{VIP}.} respectively in the Super-RDI reductions (Figure~\ref{fig:06022}) of the three epochs. We followed the procedure described in \S\ref{2M0122} to estimate systematic drift of the star behind the vortex coronagraph, with the difference of performing 2D Gaussian fits to the candidate's PSF in individual de-rotated KLIP subtracted frames (before temporal averaging) in the sequence. We do not find any significant drift of the star behind the coronagraph ($<$10 mas). The point source is stable and recovered across all reduction combinations of library size and PCs. This, along with a detection of the candidate across three epochs, significantly reduces the probability of the point source being a speckle. The candidate is detected at contrasts ($\Delta$mag) of $4.8 \pm 0.2$\,mag, $4.5 \pm 0.2$\,mag, and $4.9 \pm 0.2$\,mag with an $M_{L'}$ of $9.9 \pm 0.2$\,mag, $9.5 \pm 0.2$\,mag, $10.0 \pm 0.2$\,mag on 2016.86, 2017.04, and 2017.85 respectively. $M_{L'}$ is consistent within uncertainties across the three epochs. The ADI non-detection across all three epochs is not surprising and can be attributed to the small P.A.~rotation ($<$10$^\circ$) accrued in each of the three datasets. This leads to self-subtraction of the signal.

\begin{figure*}
    \centering
    \includegraphics[scale=0.06]{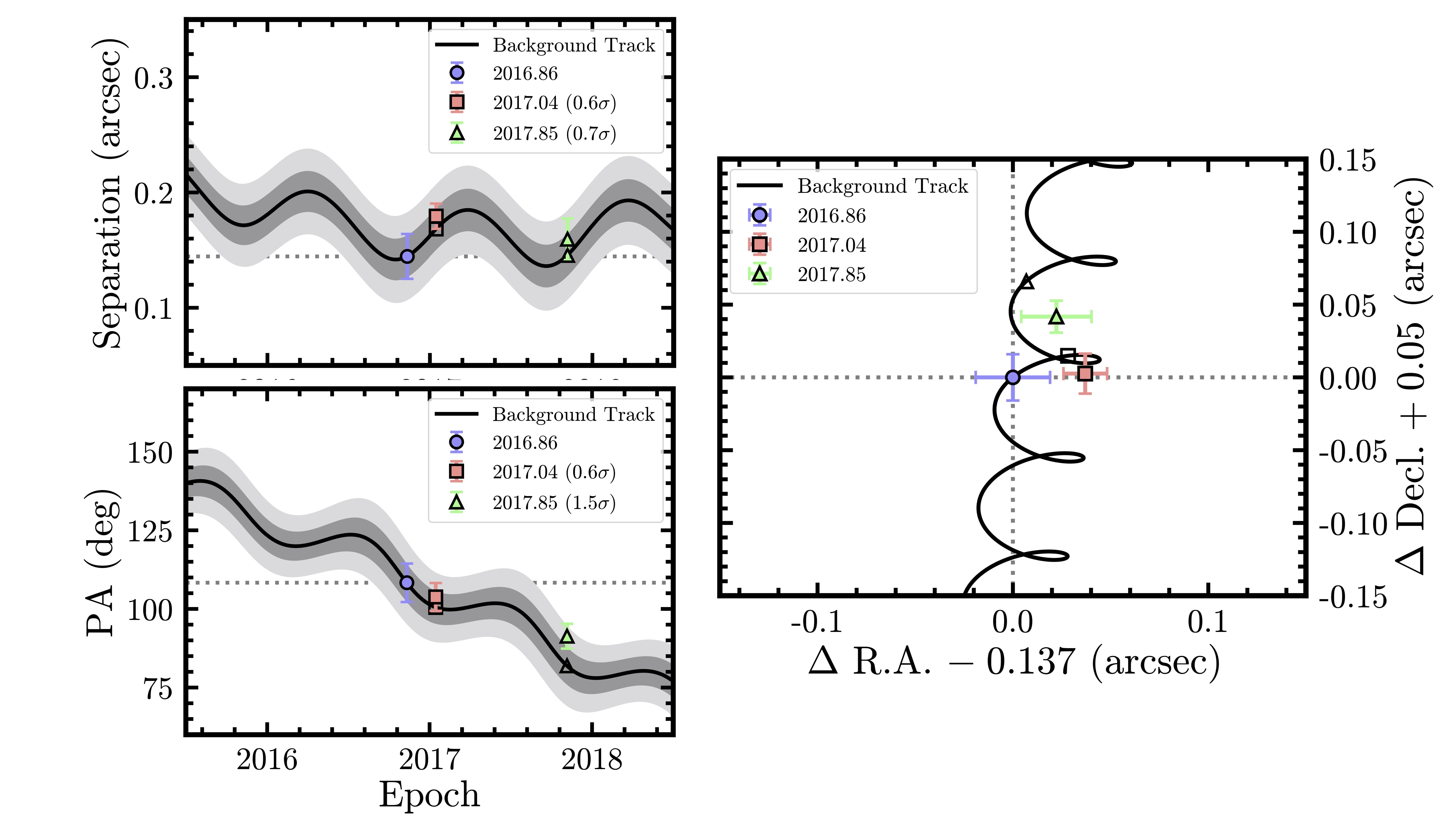}
    \caption{A common proper motion test based on three epochs of relative astrometry between 2MASS J0602-1634 and the candidate companion CC1. The left panels show the predicted relative motion of a stationary background source in separation (top; black solid line) and PA (bottom; black solid line) based on the first epoch of relative astrometry (light purple circles). Gray shaded regions represent 1$\sigma$ and 2$\sigma$ uncertainties. The predicted positions at 2017.04 and 2017.85 are plotted as an open square and an open triangle respectively. The light pink squares and green triangles represent our measurements at the respective epochs. The separation and PA measurements for the 2017.04 and 2017.85 epoch detections are consistent with the predicted background tracks within uncertainties. The right panel shows the same background track comparison but in $\Delta$R.A. and $\Delta$Decl. space instead of separation and PA.}
    \label{fig:pm}
\end{figure*}

\begin{figure*}
    \centering
    \includegraphics[scale=0.06]{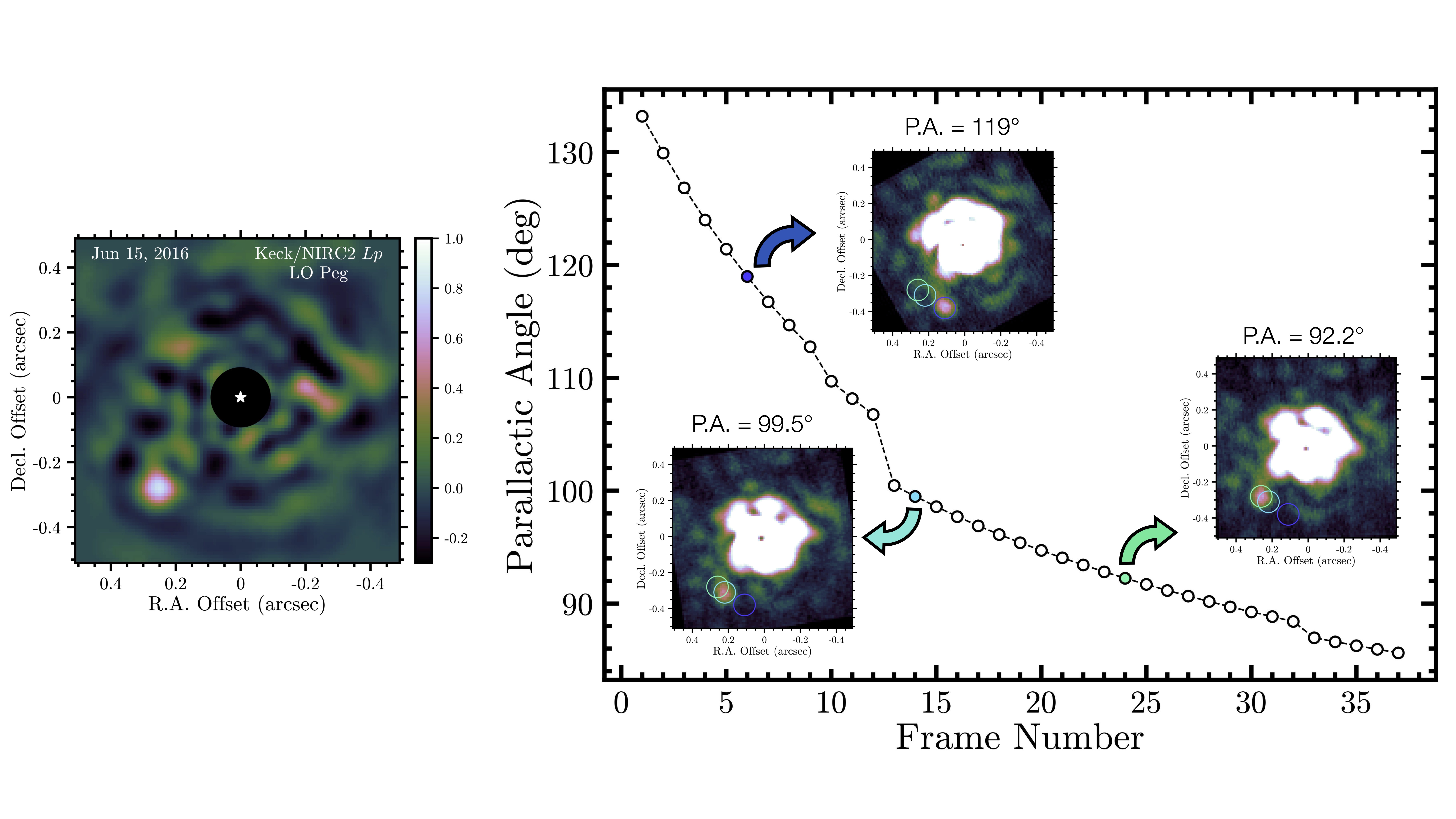}
    \caption{\emph{Left:} Final PSF-subtracted images of LO Peg obtained with Super-RDI. North is up and east is to the left. The image and color bar are depicted in linear scale in normalized analog to digital units and have been smoothed using a Gaussian kernel with a 1.5 pixel standard deviation (40\% of the width of the instrumental PSF) to average out high-frequency noise. \emph{Right:} Parallactic angle of individual frames obtained across the entire observation sequence. Presented as insets are three representative \emph{de-rotated pre-processed frames} (north is up and east is to the left) in the sequence showing that the point source in our final PSF-subtracted image rotates across the sequence as expected for a speckle.}
    \label{fig:LO}
\end{figure*}

\emph{Host star age analysis:} Based on a kinematic analysis with BANYAN-$\Sigma$, 2MASS~J06022455-1634494 is not associated with any young moving group. We must rely on spectroscopic, activity, and photometric indicators to place constraints on the age of the star. 

(1) \emph{Spectroscopic Youth Indicators.} Pre-main-sequence stars undergoing contraction exhibit weaker surface gravity compared to stars on the main sequence. The alkali doublet lines of Na~I ($\lambda$8183/8195) and K~I ($\lambda$7665/7699) are sensitive to surface gravity and can thus serve as a possible youth indicator for M dwarfs. Similarly, we can use the CaH III molecular absorption band from \citet{1995AJ....110.1838R} as a spectroscopic youth indicator. \citet{2014MNRAS.443.2561G} measure EW$_{\mathrm{KI}} = 0.970 \pm 0.050$ \AA $\;$for 2MASS~J06022455-1634494. This suggests an age $\gtrsim 100$ Myr \citep[Figure 5:][]{2014AJ....147...85R} based on the star's $V - K_S = 3.934$ \citep{2018MNRAS.475.1960F}, though \citet{2014AJ....147...85R} advise caution when interpreting the line at $V - K_S < 5$. \citet{2006AJ....132..866R} measure a CaH III index = 0.86 for 2MASS~J06022455-1634494. This index value is slightly smaller (indicating higher surface gravity) than that of M0 dwarfs in the \citet{2011AJ....141...97W} catalog \citep[Figure 14:][]{2014ApJ...794..146T}, which largely consists of older field dwarfs. 

(2) \emph{H$\alpha$ emission.} Hydrogen emission is a tracer of accretion and magnetic activity and has been associated with stellar age \citep[e.g.,][]{1963ApJ...138..832W, 1972ApJ...171..565S,  1990PASP..102..166E, 2003AJ....126.2997B}. The detection of H$\alpha$ emission can be used to place an upper limit on the age of a star. Literature measurements of EW$_{\mathrm{H}\alpha}$ for 2MASS~J06022455-1634494 lie in the range 1.46--2.2 \AA $\;$ \citep{2006AJ....132..866R, 2014MNRAS.443.2561G, 2015ApJ...798...41A}. Based on this, the M dwarf H$\alpha$ age-activity relation of \citet{2021AJ....161..277K} suggests an age $\lesssim 1.5$ Gyr (Figure 6 in the paper). 2MASS~J06022455-1634494's EW$_{\mathrm{H}\alpha}$ is also similar to those of M0 dwarfs in the Hyades \citep[$650 \pm 50$ Myr:][]{2015ApJ...807...24B} and Praesepe \citep[$750 \pm 100$ Myr:][]{2019ApJ...879..100D} open clusters \citep[Figure 5:][]{2014ApJ...795..161D}. Thus, the derived upper limit is consistent with the above result as well as the dynamical model-based activity lifetime of $0.8 \pm 0.6$ Gyr for M0 dwarfs \citep{2008AJ....135..785W}.

(3) \emph{X-ray and UV Emission.} \citet{2006AJ....132..866R} derive the ratio of the X-ray luminosity (from \emph{ROSAT}) to the bolometric luminosity for 2MASS~J06022455-1634494 $\mathrm{log}(L_{\mathrm{X}}/L_{\mathrm{bol}}) = -3.28$. Based on Figure 6 in \citet{2005ApJS..160..390P}, the above value is higher than those measured for Orion (ONC), NGC 2264, and Chamaeleon members in their sample. This would suggest an age $\lesssim 10$ Myr, which contradicts our H$\alpha$ emission-based analysis. The contradiction is resolved by noting that \citet{2015ApJ...798...41A} flag 2MASS~J06022455-1634494 as a near-ultraviolet (NUV) detection for reasons other than stellar youth. An early M dwarf in a binary system with an unresolved late M dwarf can appear NUV luminous due to the persistent activity of the late M dwarf. The system appears as a single early M dwarf because the early M dwarf dominates the continuuum emission. \citet{2015ApJ...798...41A} test for this configuration by computing the difference between the centroid of the source in a white-light image and in an H$\alpha$ image. The images are obtained from image cubes collected by \citet{2015ApJ...798...41A} using the Super-Nova Integral Field Spectrograph \citep[SNIFS:][]{2002SPIE.4836...61A, 2004SPIE.5249..146L}. If the source is a binary, a shift is expected since the unresolved late M dwarf is the source of stronger H$\alpha$ emission. A shift was observed for 2MASS~J06022455-1634494 and thus implies that the UV and X-ray measurements are unreliable for our age estimation purposes.

(4) \emph{Color-magnitude diagram.} Young, pre-main-sequence M dwarfs are still undergoing contraction and should thus have brighter absolute magnitudes than field-age objects. The Gaia DR3 photometry of 2MASS~J06022455-1634494 lines up with the main sequence and agrees with members of AB Doradus ($\approx$149 Myr) and older groups \citep[Appendix A in][]{2021AJ....161..277K}. This suggests a host star age $\gtrsim 150$ Myr.

Synthesizing all of the above information, we adopt an age estimate of 0.150--1.500 Gyr for 2MASS~J06022455-1634494. Assuming the point source is a gravitionally bound companion, based on the forward modeled photometry and the adopted age range, we estimate its mass to be in the range 60--95 $M_{\mathrm{Jup}}$ (combining the ranges derived across each of the three epochs). The inferred mass range suggests the point source is either a high mass brown dwarf or a low mass star. The properties of CC1 are summarized in Table~\ref{tab3}.

\emph{Common proper motion test:} The available data allows us to test whether the point source's astrometry is consistent with a background source or a gravitationally bound companion. Using the host star's parallax, proper motion, and ICRS coordinates (Tables~\ref{tab1} and~\ref{tab2}), we plot the predicted motion of the point source with respect to its first epoch astrometry (2016.86) assuming the stationary background source hypothesis. Figure~\ref{fig:pm} shows the result. Astrometry at the 2017.04 and 2017.85 epochs is consistent with the predicted background track within measurement uncertainties indicating that CC1 is likely a background object. However, it could be the case that the orbital motion closely mimics the background track and the differences are of the same order as our astrometric uncertainties. A fourth epoch of astrometry (now with a $\sim$7 year baseline) can resolve this ambiguity. If bound to the host star, it could potentially explain why the host star appears NUV luminous and exhibits a centroid shift in H$\alpha$ images for reasons discussed previously.

\subsection{Cautionary Case: LO Peg}
LO Peg is a young, single \citep{2005MNRAS.356.1501B, 2008MNRAS.387..237P} K3Vke ultra-fast rotator \citep{2003AJ....126.2048G, 2016MNRAS.459.3112K} exhibiting strong flaring activity based on H$\alpha$ and He \RomanNumeralCaps{1} D3 observations \citep{1994MNRAS.270..153J, 1999A&A...341..527E}. It is a AB Doradus young moving group \citep[$149_{-19}^{+51}$ Myr;][]{2015MNRAS.454..593B} member \citep{2004ARA&A..42..685Z}, with a BANYAN-$\Sigma$ probability of 99.7\% \citep{2018ApJ...856...23G} computed using its parallax, proper motion, and radial velocity (Table~\ref{tab2}). 

We observed LO~Peg on 2016~June~15~UT and detected a point source at $\approx$362 mas and $\approx$136.3$^\circ$ at a S/N of $\approx$11 (6.6$\sigma$ detection) in our Super-RDI reductions (Figure~\ref{fig:LO}, left). The point source is stable and recovered across all reduction combinations of library size and PCs. Based on the PSF-subtracted images, the point source appeared to be a promising substellar candidate (using the forward modeled photometry and the AB Doradus young moving group age one would find a mass in the range of 13--21 $M_{\mathrm{Jup}}$) for follow-up observations.

However, visual inspection of the pre-processed coronagraphic frames revealed a bright source at similar separations. We de-rotated the pre-processed coronagraphic frames to align them north-up based on their parallactic angles. True astrophysical sources would remain stationary in the individual de-rotated frames. We found that the position angle of the bright source changes in the de-rotated frames across the sequence providing clear evidence that it is a speckle (Figure~\ref{fig:LO}, right). Typically, we expect the signal of such a speckle to be smeared in the final PSF-subtracted image due to its changing position angle across the de-rotated frames. However, this is not seen in our reduced image likely because we acquired a majority of the frames in the second half of the sequence, where the P.A.~change was much slower than in the first half (Figure~\ref{fig:LO}, right). 

This example shows that even with the availability of a large reference library it can be challenging to fit for and eliminate atypical high spatial frequency speckles in the data. In such cases, there is value in analyzing the pre-processed coronagraphic frames or the individual PSF-subtracted frames before de-rotation to assess the nature of candidate point sources. The observed lack of smearing of the speckle also suggests that RDI may have a higher false positive rate for vertical angle mode observations when targets are scheduled asymmetrically about transit.  

\section{Conclusion}
\label{sec:con}
This work describes the Super-RDI framework, designed to improve the performance of the RDI strategy when working with large reference libraries. We developed and applied this framework in the context of Keck/NIRC2 high-contrast $L'$ imaging observations with the vortex coronagraph. Our primary results are summarized below:
\begin{enumerate}
    \item We presented a set of 288 new $L'$ observations (central wavelength of 3.776 $\mu$m) of 237 unique targets observed between 2015-12-26 and 2019-01-09. The sample was comprised of targets that were observed independently as part of two survey programs with Keck/NIRC2: the young M-star Survey (195 observations of 157 unique targets) and the Taurus Survey (93 observations of 80 unique targets). The complete set of observations consisted of 7060 image frames.
    
    \item The Super-RDI framework (1) used image similarity metrics to rank and select the best matching reference stars uniquely for each target; (2) optimized free parameters (metric, library size, number of KLIP PCs) using synthetic companion injection-recovery tests to maximize detection sensitivity; and (3) uniformly processed and conducted a sensitivity analysis of the target sample using the optimized set of reduction parameters.
    
    \item For our dataset, synthetic companion injection-recovery tests revealed that frame selection with the mean-squared error (MSE) metric combined with KLIP-based PSF subtraction using 1000--3000 frames and $<$500 principal components yields the highest average S/N for injected synthetic companions.
    
    \item We assembled Super-RDI and ADI principal contrast curves for each target by selecting the deepest $5\sigma$ contrast value at each separation across all reduction parameters. The difference between the Super-RDI and ADI contrasts averaged across all targets in our sample (with a typical P.A.~rotation of $\sim$10$^\circ$) showed that Super-RDI improves upon a widely used implementation of ADI-based PSF subtraction at angular separations $<0\farcs4$. Super-RDI gained as much as 0.25\,mag in contrast at $0\farcs25$ and 0.4\,mag in contrast at $0\farcs15$. Our results were an improvement over traditional RDI observations with Keck/NIRC2 \citep{2018AJ....156..156X} in separation space. For best performance with a similar observational design, we recommended complementing Super-RDI with dedicated reference star observations for small angular separation ($\lesssim 0\farcs4$) point source detection with Keck/NIRC2. These results also showed that ADI is the preferred strategy for wide separation ($\gtrsim 0\farcs4$) companion searches in the young M-star survey. 
    
    \item Based on a comparison of Super-RDI and ADI's performance as a function of P.A.~rotation, we found that the separation space over which Super-RDI performs better than ADI shrinks as the P.A.~rotation of the observations increases. The results showed that ADI is the preferred strategy at small angular separations for individual observations where larger P.A.~rotations (generally $>$30$^\circ$) can be accumulated.
    
    \item We investigated the performance of Super-RDI as a function of host star \emph{W1} and \emph{R} magnitude. Negative slopes are found for best-fit lines to Super-RDI contrast at $0\farcs2$ and $0\farcs4$ as a function of both \emph{W1} and \emph{R} magnitude. This indicated decreasing Super-RDI performance for fainter targets in the sample. Additionally, we found that the performance of Super-RDI plateaus for stars brighter than 6\,mag in \emph{W1} and 9.5\,mag in \emph{R}.
    
    \item We analyzed the difference in contrasts (at $0\farcs2$ and $0\farcs4$) achieved by Super-RDI and ADI as a function of host star \emph{W1} and \emph{R} magnitude. It was found that Super-RDI shows an improvement over ADI for stars in our sample fainter than 6\,mag in \emph{W1} and 9.5\,mag in \emph{R} at both separations. ADI achieves deeper contrasts at both separations at all other magnitudes.
    
    \item We directly imaged a point source around 2MASS~J06022455-1634494 at contrasts of 4.5--5.4\,mag (S/N $\approx$5) in three observation epochs at $\approx$163 mas ($\approx$6.6 au) and $\approx$101.4$^\circ$. 2MASS~J06022455-1634494 is a field M0 star with an estimated age of 0.150--1.500 Gyr. A common proper motion test indicated that the point source is likely to be a background object. 

    \item We provided pre-processed $L'$ coronagraphic frames, associated parallactic angles, unocculted PSF frames, Super-RDI reduced images, and detection limits for all observations in the young M-star survey on Zenodo\footnote{\url{https://zenodo.org/records/12747613}} for public access.

\end{enumerate}

Super-RDI is a promising framework for the implementation of RDI with large reference libraries to Keck/NIRC2 high-contrast imaging surveys targeting companions at small angular separations. The technique provides improvements over traditional RDI observations in separation space and mitigates the limitations of ADI such as scheduling difficulty, constrained sky coverage, and self-subtraction at small separations. Combining gains in post-processing with instrumental upgrades in the areas of AO and coronagraphy to improve sensitivity at angular separations $<$~0\farcs5 will be the key to unlocking new parameter spaces for giant planet demographic studies. The recent direct imaging detections of 8.2~$M_{\mathrm{Jup}}$ $\beta$~Pictoris~c at 0\farcs13 \citep{2020A&A...642L...2N}, 24~$M_{\mathrm{Jup}}$ HIP~21152~B at 0\farcs37 \citep{2022ApJ...934L..18K, 2023AJ....165...39F}, 12.7~$M_{\mathrm{Jup}}$ HD~206893~c at 0\farcs11 \citep{2023A&A...671L...5H}, 16.1~$M_{\mathrm{Jup}}$ HIP~99770~b at 0\farcs43 \citep{2023Sci...380..198C}, and 3.2~$M_{\mathrm{Jup}}$ AF~Lep~b at 0\farcs34 \citep{2023A&A...672A..94D, 2023ApJ...950L..19F, 2023A&A...672A..93M} represent exciting steps in this direction. More broadly, for the nearest and youngest stars, this will facilitate building a larger direct imaging sample of the most typical giant planets at physical separations of 1--10 au for population-level orbital and atmospheric characterization studies. 

\section*{Acknowledgements}
The authors would like to thank the anonymous referee for many helpful comments which improved the manuscript. We also thank Michael C. Liu for helpful discussions on M dwarf age-dating techniques. A.S. acknowledges support from the Summer Undergraduate Research Fellowship (SURF) from California Institute of Technology. B.P.B. acknowledges support from the National Science Foundation grant AST-1909209, NASA Exoplanet Research Program grant 20-XRP20$\_$2-0119, and the Alfred P. Sloan Foundation. This research is partially supported by NASA ROSES XRP, award 80NSSC19K0294 and by the Gordon and Betty Moore Foundation through grant GBMF8550. This project has received funding from the European Research Council (ERC) under the European Union's Seventh Framework Program (grant agreement No 337569) and the European Union's Horizon 2020 research and innovation programme (grant agreement No 819155), and from the Wallonia-Brussels Federation (grant for Concerted Research Actions). O.A.\ is a Senior Research Associate of the Fonds de la Recherche Scientifique - FNRS. The Infrared Pyramid Wavefront Sensor at W. M. Keck Observatory was developed with support from the National Science Foundation under grants AST-1611623 and AST-1106391, as well as the Heising Simons Foundation under the Keck Planet Imager and Characterizer project. The data presented herein were obtained at the W. M. Keck Observatory, which is operated as a scientific partnership among the California Institute of Technology, the University of California and the National Aeronautics and Space Administration. The Observatory was made possible by the generous financial support of the W. M. Keck Foundation. The authors wish to recognize and acknowledge the very significant cultural role and reverence that the summit of Maunakea has always had within the indigenous Hawaiian community. We are most fortunate to have the opportunity to conduct observations from this mountain. The computations presented here were conducted in the Resnick High Performance Center, a facility supported by Resnick Sustainability Institute at the California Institute of Technology. Part of this work was carried out at the Jet Propulsion Laboratory, California Institute of Technology, under contract with NASA. This research has made use of the Keck Observatory Archive (KOA), which is operated by the W. M. Keck Observatory and the NASA Exoplanet Science Institute (NExScI), under contract with the National Aeronautics and Space Administration. This research has made use of the VizieR catalog access tool, CDS, Strasbourg, France (DOI:10.26093/cds/vizier). The original description of the VizieR service was published in \citet{2000A&AS..143...23O}. This publication also makes use of data products from the Wide-field Infrared Survey Explorer, which is a joint project of the University of California, Los Angeles, and the Jet Propulsion Laboratory/California Institute of Technology, funded by the National Aeronautics and Space Administration; and the Spanish Virtual Observatory (https://svo.cab.inta-csic.es) project funded by MCIN/AEI/10.13039/501100011033/ through grant PID2020-112949GB-I00. 

\emph{Facility:} Keck II (NIRC2).

\emph{Additional Software/Resources:}, \texttt{Astropy} \citep{2013A&A...558A..33A, 2018AJ....156..123A}, \texttt{Matplotlib} \citep{2007CSE.....9...90H}, \texttt{NumPy} \citep{harris2020array}, \texttt{Photutils} \citep{2019zndo...3568287B}, \texttt{scikit-image} \citep{2014arXiv1407.6245V}, \texttt{VIP} \citep{2017AJ....154....7G, Christiaens2023}, and the NASA Astrophysics Data System (ADS).

\appendix
\restartappendixnumbering
\section{ADI Optimization for Young M-star Survey Targets}
\label{app}
The ADI reduction procedure for our young M-star survey sample is described in \S\ref{sec:adi}. For the $101\times101$ pixels frame size reductions with a 1 FWHM central mask (most relevant for our comparisons with Super-RDI), we conducted both full frame and annular PCA-KLIP reductions for different number of principal components. For annular PCA-KLIP reduction, we additionally employed a rotation gap criterion. Due to the low P.A.~rotation of our dataset, the 0.1, 0.5, 1, and 2 FWHM rotation gap criterion could only be applied for 121, 4, 2, and 1 observation (out of 195) respectively in our sample without excluding all available frames. It is also important to note that the results presented here are specific to our sample and cannot be broadly generalized to other ADI datasets, for example, those with larger P.A.~rotation.

\begin{figure*}
    \centering
    \includegraphics[scale=0.06]{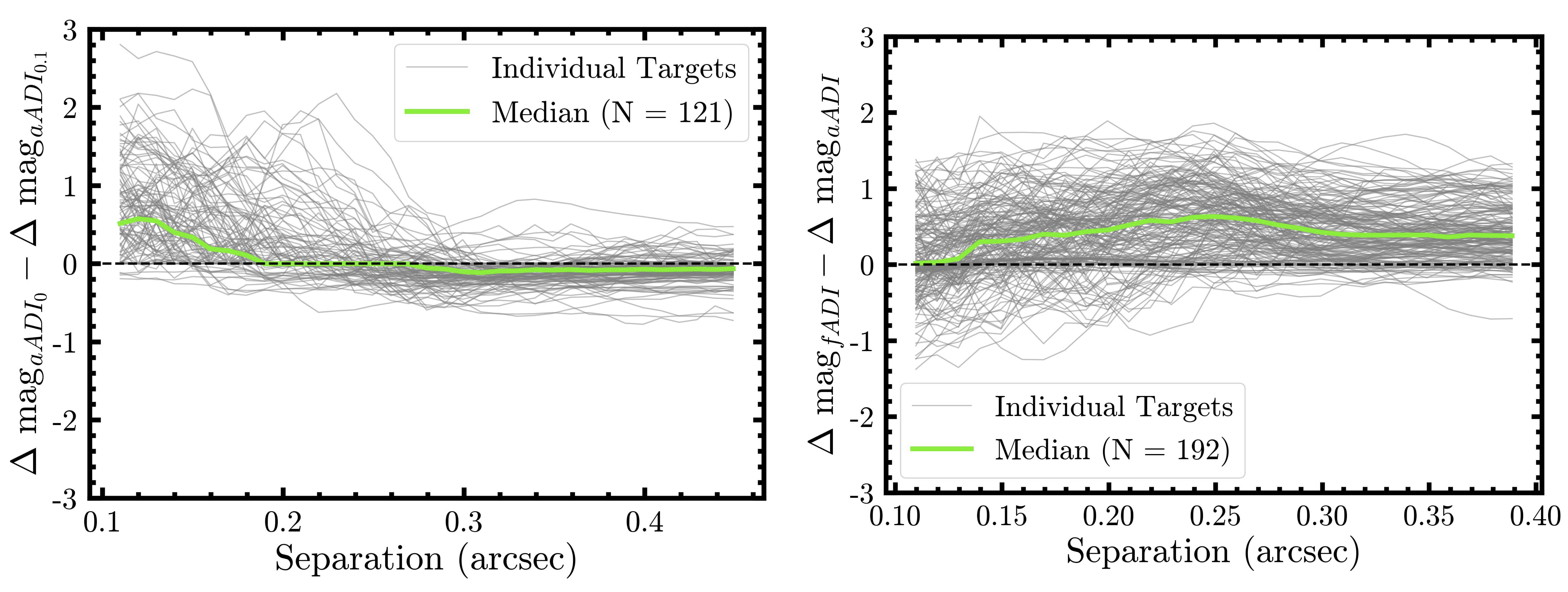}
    \caption{\emph{Left:} Difference in contrast (units of $\Delta$mag) achieved between annular ADI with a rotation gap criterion of 0.1 FWHM ($aADI_{0.1}$) and annular ADI with no rotation gap criterion ($aADI_0$) as a function of separation (units of arcseconds). A black dashed line marks zero difference and corresponds to both strategies achieving the same sensitivity level. Grey curves correspond to individual targets in the sample and the green curve corresponds to the median computed across all targets. \emph{Right:} Same as the left panel but presenting the difference in contrast achieved between full frame ADI ($fADI$) and annular ADI ($aADI$). The best performing strategy varies depending on the individual target.}
    \label{fig:adiopt}
\end{figure*}

We compare the contrast achieved by annular PCA-KLIP reductions without a rotation gap criterion and those with a rotation gap criterion of 0.1 FWHM (left panel of Figure~\ref{fig:adiopt}). Contrast curves generated with varying PCs in each case are combined together by selecting the deepest contrast value at each separation to generate a ``principal" contrast curve for the two reductions of each target. The reduction strategy that achieves deeper contrast varies significantly between individual targets as well as separation from the host star. Generally, we find that annular PCA-KLIP with a rotation gap criterion of 0.1 FWHM provides better performance at separations $\gtrsim0\farcs25$ for our dataset. We also compare the contrast achieved by full frame and annular PCA-KLIP reductions (right panel of Figure~\ref{fig:adiopt}). Contrast curves generated with varying PCs (for full frame and annular reductions) and with varying values of rotation gap criteria (for annular reductions) are combined together by selecting the deepest contrast value at each separation to generate a ``principal" contrast curve for the full frame and annular PCA-KLIP reduction of each target. Based on the principal contrast curves, the reduction strategy that achieves deeper contrast depends on the individual target. However, for the majority of targets in our sample, we find that full frame PCA-KLIP is preferred over annular PCA-KLIP.

\section{Table of Optimal Contrast Curves}
\label{app:B}

We provide optimal contrasts at select separations for targets in the young M-star survey in Table~\ref{tabA1}.
\input{tabA1}

\bibliography{\string Sanghi_Astro_Bib.bib}

\end{document}

%% file: tab1.tex
\begin{longrotatetable}
\begin{deluxetable*}{cccccccccc}
\centering
\movetabledown=3mm
\tabletypesize{\scriptsize}
\label{tab1}
\tablecaption{Keck/NIRC2 $L'$ Vortex Coronagraphic Observations}
\tablehead{\colhead{Name} & \colhead{$\alpha_{2000.0}$} & \colhead{$\delta_{2000.0}$} & \colhead{UT Date} & \colhead{$N\times\mathrm{Coadds}\times t_{\mathrm{exp}}$} & \colhead{$\Delta$P.A.} & \colhead{Seeing} & \colhead{\emph{W1}} & \colhead{\emph{R}} & \colhead{\emph{R} Ref. Code} \\ 
\colhead{} & \colhead{(hh:mm:ss.ssss)} & \colhead{($\pm$dd:mm:ss.sss)} & \colhead{} & \colhead{(s)} & \colhead{($^\circ$)} & \colhead{(arcsec)} & \colhead{(mag)} & \colhead{(mag)} & \colhead{}}

\startdata
\cutinhead{Young M-star Survey Targets}                                                                                                
2MASS J00243202-2522528 & 00:24:32.0175 & -25:22:52.952 & 2016-07-18 & $18\times45\times1.0$ & 7.2 & $0.52\pm0.13$ & $8.800\pm0.024$ & $13.221\pm0.130$ & 1 \\ 
2MASS J00275023-3233060 & 00:27:50.2331 & -32:33:06.306 & 2016-11-05 & $20\times45\times1.0$ & 7.0 & $0.82\pm0.22$ & $7.782\pm0.028$ & $11.790\pm0.030$ & 2 \\ 
2MASS J00275023-3233060 & 00:27:50.2331 & -32:33:06.306 & 2016-09-11 & $20\times45\times1.0$ & 6.9 & $0.56\pm0.18$ & $7.782\pm0.028$ & $11.790\pm0.030$ & 2 \\ 
GJ2006B & 00:27:50.3535 & -32:33:24.130 & 2016-09-12 & $14\times45\times1.0$ & 4.8 & $0.65\pm0.22$ & $7.931\pm0.025$ & $12.040\pm0.020$ & 2 \\ 
$[$SLS2012$]$ PYC J00390+1330 & 00:39:03.4171 & +13:30:16.848 & 2018-07-30 & $23\times45\times1.0$ & 52.6 & $0.54\pm0.21$ & $9.843\pm0.023$ & \nodata & \nodata \\ 
2MASS J00551501+3015156 & 00:55:15.0287 & +30:15:15.567 & 2016-09-23 & $27\times45\times1.0$ & 35.4 & $0.4$ & $9.105\pm0.023$ & $12.390$ & 3 \\ 
2MASS J00551501+3015156 & 00:55:15.0287 & +30:15:15.567 & 2016-09-11 & $20\times45\times1.0$ & 26.1 & $0.56\pm0.18$ & $9.105\pm0.023$ & $12.390$ & 3 \\ 
2MASS J01123504+1703557 & 01:12:35.0523 & +17:03:55.571 & 2016-10-17 & $20\times45\times1.0$ & 24.6 & $0.93\pm0.24$ & $9.256\pm0.022$ & $13.780\pm0.090$ & 1 \\
\enddata

\tablerefs{(1) \citet{2012yCat.1322....0Z}; (2) \citet{2014AJ....147...85R}; (3) \citet{2007AA...467..785N}; (4) \citet{2007ApJS..168..128B}; (5) \citet{2005AJ....129.1483L}; (6) \citet{2009yCat.1315....0Z}; (7) \citet{2010MNRAS.403.1949K}; (8) \citet{2009AJ....137.4186L}; (9) \citet{2003yCat.1289....0Z}; (10) \citet{2007ApJ...658..480M}; (11) \citet{2007AJ....133.2825R}; (12) \citet{2003AJ....125..984M}; (13) \citet{2008MNRAS.389..585C}; (14) \citet{2002yCat.2237....0D}; (15) \citet{2008AJ....135..850D}; (16) \citet{2006AJ....132.2469K}; (17) \citet{2012AA...543A..59M}; (18) \citet{2020ApJS..247...44A}; (19) \citet{2008ApJ...689..513T}; (20) \citet{2012ApJ...750..157C}; (21) \citet{2007AA...467..785N}.}

\tablenotetext{a}{\emph{W1} magnitude is retrieved from \citet{2012yCat.2311....0C}.}

\tablenotetext{b}{The Taurus survey targets only serve as reference stars for Super-RDI reductions of  the young M-star survey targets. PSF subtraction of the Taurus survey targets will be presented in Bryan et al. (in preparation).}

\tablecomments{The mean and standard deviation of seeing measurements for the night are presented as recorded by the DIMM Seeing Monitor. Seeing measurements quoted without uncertainties correspond to approximate values measured by the CFHT WX Tower Seeing Monitor. \emph{W1} magnitudes are retrieved from \citet{2014yCat.2328....0C} unless otherwise noted next to the measurement. \emph{R} magnitude uncertainties are not quoted for objects where the corresponding SIMBAD reference does not provide one.}

\end{deluxetable*}
\end{longrotatetable}

%% file: tab2.tex
\centerwidetable
\begin{deluxetable*}{cccccc}
\label{tab2}
\centering
\tabletypesize{\small}
\tablecaption{BANYAN-$\Sigma$ Inputs for Host Star Membership Analysis}
\tablehead{\colhead{Name} & \colhead{$\pi$} & \colhead{$\mu_{\alpha}$\tablenotemark{\scriptsize a}} & \colhead{$\mu_{\delta}$} & \colhead{$v\:\mathrm{sin}\:i$} & \colhead{References} \\ \colhead{} & \colhead{(mas)} & \colhead{(mas/yr)} & \colhead{(mas/yr)} & \colhead{(km/s)} & \colhead{}}

\startdata
2MASS J01225093-2439505 & $29.6409 \pm 0.0273$ & $120.215 \pm 0.034$ & $-123.561 \pm 0.021$ & $11.4 \pm 0.2$ & 1, 1, 1, 4 \\
2MASS J23513366+3127229 & $23.1248 \pm 0.0217$ & $106.656 \pm 0.021$ & $-87.886 \pm 0.016$ & $-13.55 \pm 0.07$ & 1, 1, 1, 2\\
2MASS J06022455-1634494 & $24.8661 \pm 0.4167$ & $-8.222 \pm 0.368$ & $-67.489 \pm 0.365$ & $-8.22 \pm 0.04$ & 1, 1, 1, 2\\
LO Peg & $41.2912 \pm 0.0169$ & $134.654 \pm 0.013$ & $-144.889 \pm 0.008$ & $-4.44 \pm 0.01$ & 1, 1, 1, 3\\
\enddata

\tablenotetext{a}{Proper motion in R.A. includes a factor of cos$\:\delta$.}
\tablerefs{(1) \citet{2020yCat.1350....0G}; (2) \citet{2018MNRAS.475.1960F}; (3) \citet{2021AA...645A..30Z}; (4) \citet{2014ApJ...788...81M}.}
\end{deluxetable*}

%% file: tab3.tex
\centerwidetable
\begin{deluxetable*}{cccccccccc}
\label{tab3}
\centering
\tabletypesize{\scriptsize}
\tablecaption{Properties of Point Source Detections}
\tablehead{\colhead{Name} & \colhead{UT Date} & \colhead{Angular Separation ($\rho$)} & \colhead{P.A. ($\theta$)} & \colhead{$\pi$} & \colhead{Physical Separation} & \colhead{$\Delta$mag} & \colhead{$M_{Lp}$} & \colhead{Age} & \colhead{Inferred Mass} \\ \colhead{} & \colhead{} & \colhead{(arcsec)} & \colhead{($^\circ$)} &  \colhead{(mas)} & \colhead{(au)} & \colhead{(mag)} & \colhead{(mag)} & \colhead{(Myr)} & \colhead{($M_{\mathrm{Jup}}$)} }

\startdata
2MASS J01225093-2439505 B & 2016-10-17 & $1.458 \pm 0.053$ & $217.0 \pm 2.7$ & $29.6409 \pm 0.0273$ & $49.2 \pm 0.4$ & $4.4 \pm 0.1$ & $10.8 \pm 0.1$ & $149^{+51}_{-19}$ & 18--28 \\ 
 & &  & &  &  &  &  \\ 
2MASS J23513366+3127229 B & 2017-09-09 & $2.393 \pm 0.027$ & $90.6 \pm 1.0$ & $23.1248 \pm 0.0217$ & $103.5 \pm 0.5$ & $4.3 \pm 0.1$ & $10.0 \pm 0.1$ & $149^{+51}_{-19}$ & 36--42 \\ 
 & 2018-07-30 & $2.427 \pm 0.049$ & $90.1 \pm 0.9$ & $23.1248 \pm 0.0217$ & $105.0 \pm 0.5$ & $4.4 \pm 0.1$ & $10.0 \pm 0.1$ & $149^{+51}_{-19}$ & 36--42 \\
 & &  & &  &  &  &  \\ 
CC1\tablenotemark{\scriptsize a} & 2016-11-11 & $0.145 \pm 0.019$ & $108.3 \pm 6.1$ & $24.8661 \pm 0.4167$ & $5.8 \pm 0.8$ & $4.8 \pm 0.2$ & $9.9 \pm 0.2$ & 150--1500\tablenotemark{\scriptsize b} & 62--86 \\ 
 & 2017-01-15 & $0.179 \pm 0.011$ & $103.8 \pm 4.4$ & $24.8661 \pm 0.4167$ & $7.2 \pm 0.4$ & $4.5 \pm 0.2$ & $9.5 \pm 0.2$ & 150--1500\tablenotemark{\scriptsize b} & 75--95 \\ 
 & 2017-11-06 & $0.160 \pm 0.018$ & $91.3 \pm 3.9$ & $24.8661 \pm 0.4167$ & $6.4 \pm 0.6$ & $4.9 \pm 0.2$ & $10.0 \pm 0.2$ & 150--1500\tablenotemark{\scriptsize b} & 60--84 \\
\enddata

\tablenotetext{a}{This point source detection is a likely background star. In this table, we present the age of the primary star and a mass range for the point source assuming it is gravitationally bound to 2MASS J06022455-1634494.}
\tablenotetext{b}{2MASS J06022455-1634494's age is determined based on spectroscopic, activity, and photometric indicators.}
\tablecomments{Parallaxes are obtained from \emph{Gaia} \citep{2020yCat.1350....0G}. Ages are obtained from \citet{2015MNRAS.454..593B} unless noted otherwise.}
\end{deluxetable*}

%% file: tabA1.tex
\begin{longrotatetable}
\begin{deluxetable*}{cccccccccccccccccc}
\centering
\movetabledown=5mm
\tabletypesize{\scriptsize}
\label{tabA1}
\tablecaption{Keck/NIRC2 $L'$ Optimal Contrast for Young M-star Survey Targets at Select Separations}
\tablehead{\colhead{Name} & \colhead{UT Date} & \colhead{0\farcs1} & \colhead{0\farcs2} & \colhead{0\farcs3} & \colhead{0\farcs4} & \colhead{0\farcs5} & \colhead{0\farcs6} & \colhead{0\farcs7} & \colhead{0\farcs8} & \colhead{0\farcs9} & \colhead{1\farcs0} & \colhead{1\farcs25} & \colhead{1\farcs5} & \colhead{1\farcs75} & \colhead{2\farcs0} & \colhead{2\farcs25} & \colhead{2\farcs5} \\ 
\colhead{} & \colhead{} & \colhead{($\Delta$mag)} & \colhead{($\Delta$mag)} & \colhead{($\Delta$mag)} & \colhead{($\Delta$mag)} & \colhead{($\Delta$mag)} & \colhead{($\Delta$mag)} & \colhead{($\Delta$mag)} & \colhead{($\Delta$mag)} & \colhead{($\Delta$mag)} & \colhead{($\Delta$mag)} & \colhead{($\Delta$mag)} & \colhead{($\Delta$mag)} & \colhead{($\Delta$mag)} & \colhead{($\Delta$mag)} & \colhead{($\Delta$mag)} & \colhead{($\Delta$mag)}}

\startdata
2MASS J00243202-2522528 & 2016-07-18 & 4.14 & 5.40 & 6.05 & 6.77 & 7.25 & 7.01 & 6.90 & 7.33 & 7.28 & 7.49 & 7.87 & 7.99 & 7.93 & 8.01 & 8.06 & 8.08 \\
2MASS J00275023-3233060 & 2016-11-05 & 4.50 & 5.83 & 7.20 & 7.29 & 7.52 & 7.70 & 8.05 & 8.28 & 8.59 & 8.60 & 8.76 & 8.89 & 9.08 & 9.17 & 9.22 & 9.22 \\
2MASS J00275023-3233060 & 2016-09-11 & 3.49 & 4.89 & 6.57 & 6.77 & 6.57 & 6.71 & 7.49 & 7.75 & 7.86 & 7.67 & 8.08 & 8.06 & 8.32 & 8.27 & 8.19 & 8.57 \\
GJ2006B & 2016-09-12 & 4.25 & 5.90 & 6.77 & 7.01 & 7.16 & 7.41 & 7.84 & 7.89 & 7.81 & 7.74 & 7.88 & 7.99 & 8.01 & 8.19 & 8.20 & 8.24 \\
2MASS J00551501+3015156 & 2016-09-23 & 3.98 & 5.50 & 6.64 & 7.00 & 6.93 & 7.28 & 7.42 & 7.36 & 7.40 & 7.43 & 7.53 & 7.53 & 7.54 & 7.74 & 7.81 & 7.80 \\
2MASS J00551501+3015156 & 2016-09-11 & 3.83 & 5.93 & 6.19 & 6.02 & 6.03 & 6.37 & 6.87 & 6.87 & 7.06 & 7.23 & 7.32 & 7.43 & 7.60 & 7.66 & 7.68 & 7.60 \\
2MASS J01123504+1703557 & 2016-10-17 & 3.24 & 4.98 & 5.98 & 6.32 & 6.75 & 7.12 & 7.12 & 7.26 & 7.35 & 7.38 & 7.43 & 7.46 & 7.51 & 7.59 & 7.66 & 7.73 \\
2MASS J01220441-3337036 & 2015-12-27 & 3.51 & 5.61 & 6.84 & 7.03 & 7.14 & 7.39 & 7.84 & 8.05 & 8.37 & 8.44 & 8.69 & 8.80 & 8.91 & 8.94 & 9.02 & 8.99 \\
\enddata

\end{deluxetable*}
\end{longrotatetable}